\documentclass[12pt]{article}
\usepackage[colorlinks,linkcolor=blue,citecolor=blue,urlcolor=blue,bookmarks,bookmarksnumbered]{hyperref}
\usepackage{mathpazo,charter}
\usepackage[scaled=0.9]{helvet}
\usepackage{XXXE}
\usepackage{graphicx,color}
\usepackage{booktabs,multirow}

\def\Dt#1{\accentset{\hbox{.}}{#1}}
\def\dt#1{\accentset{\hbox{\small.}}{#1}}	% dot-over for sp/sb
\def\cb#1#2{\setlength\fboxsep{1pt}\colorbox{#1}{\color{#1}\fbox{\color{black}#2}}}
\def\cB#1{\hbox to0pt{\setlength\fboxsep{0pt}\setlength\fboxrule{.4pt}\hss\color{grey3}%
           \fbox{\cb{white}{#1}}\hss}}

\def\Cx#1{\makebox[0pt][c]{#1}}
\def\bc{\text{\bsf\slshape c}}
\def\Bc{{\textbf{C}}}
\def\Lx#1{\makebox[0pt][l]{#1}}
\def\Rx#1{\makebox[0pt][r]{#1}}
\def\CP{\mathop\textsl{CP}\nolimits}
\def\sC{\mathscr{C}}
\def\fA{\textsf{\bfseries A}}
\let\bs=\boldsymbol

\def\rD{{\rm D}}
\def\bD{\boldsymbol{D}}
\def\bDb{\hbox{\kern1pt\vrule height10pt depth-9.2pt width6pt\kern-8pt{$\boldsymbol D$}}}

\def\sR{\mathscr{R}}
\def\BRST{\mathop{\mathcal{S}\mkern-1mu_{\sss\text{B}}}}
\def\Spin{\mathop\textsl{Spin}\nolimits}

\def\vC#1{\vcenter{\hbox{\hss#1\hss}}}
\def\vdt{\partial_\tau}
\def\pp{{=\mkern-10mu|\mkern4mu}} % or \def\pp{{++}}
\def\mm{{=}}                                % or \def\mm{{--}}
\definecolor{Green}  {rgb}{0.10,0.70,0.10} %  1
\definecolor{Orange} {rgb}{1.00,0.50,0.15} %  2
\definecolor{Red}    {rgb}{0.90,0.00,0.12} %  3
\definecolor{Purple} {rgb}{0.42,0.15,0.45} %  4
\definecolor{Turque} {rgb}{0.00,0.65,0.85} %  5
\definecolor{Blue}   {rgb}{0.00,0.00,1.00} %  6
\definecolor{Magenta}{rgb}{1.00,0.00,1.00} %  7
\definecolor{Gold}   {rgb}{1.00,0.75,0.25} %  8
\definecolor{Seaweed}{rgb}{0.01,0.24,0.09} %  9
\definecolor{Brown}  {rgb}{0.43,0.26,0.32} % 10
\definecolor{grey1}  {rgb}{0.20,0.20,0.20} % 11
\definecolor{grey2}  {rgb}{0.40,0.40,0.40} % 12
\definecolor{grey3}  {rgb}{0.60,0.60,0.60} % 13
\definecolor{grey4}  {rgb}{0.80,0.80,0.80} % 14
\definecolor{grey5}  {rgb}{0.90,0.90,0.90} % 15
\definecolor{Teal}   {rgb}{0.00,0.30,0.55} % 16
\definecolor{Hey}    {rgb}{0.90,0.05,0.40} % 17
\definecolor{plum}   {rgb}{0.40,0.00,0.60} % 18
\def\C#1#2{{\ifcase#1\or
             \color{Green}\or \color{Orange}\or \color{Red}\or
              \color{Purple}\or \color{Turque}\or \color{Blue}\or
               \color{Magenta}\or \color{Gold}\or \color{Seaweed}\or
                \color{Brown}\or\color{grey1}\or\color{grey2}\or
                 \color{grey3}\or\color{grey4}\or\color{grey5}\else
                  \color{Teal}\fi#2}}

\makeatletter
\newcounter{comNum}\@addtoreset{comNum}{section}
               {\par\medskip\leftskip=2pc\noindent\small%
                 \baselineskip=11pt plus1pt minus.5pt%
                 \abovedisplayskip=.5em plus1pt minus.5pt
                 \belowdisplayskip=.8em plus1em minus.5pt
                 \abovedisplayshortskip=.15em plus.5pt
                 \belowdisplayshortskip=.25em plus.5pt
                 \addtocounter{comNum}{1}%
                  \def\@currentlabel{{\thesection.\arabic{comNum}}}%
                  \color{Teal}{{\sf\bfseries Comment\kern1mm%
                                 \thesection.\arabic{comNum}}:}~%
                   \label{#1}\ignorespaces}
               {\par\medskip}
\makeatother

 \SfTitles %section titles more modest
 \allowdisplaybreaks
\begin{document}

 \begin{center}
{\LARGE\sf\bfseries\boldmath
  On Supermultiplet Twisting and Spin-Statistics
 }\\*[5mm]
  %\vfill
{\large\sf\bfseries 
              T.~H\"{u}bsch
}\\*[1mm]
{\small\it
      Department of Physics \&\ Astronomy, Howard University, Washington, DC\\[-1mm]
      Department of Physics, University of Central Florida, Orlando FL
  \\[-4pt] {\tt\slshape  thubsch\,@\,howard.edu}
 }\\[5mm]
  %\vfill
{\sf\bfseries ABSTRACT}\\[3mm]
\parbox{154mm}{\addtolength{\baselineskip}{-2pt}\parindent=2pc\noindent
Twisting of off-shell supermultiplets in models with $1{+}1$-dimensional spacetime has been discovered in 1984, and was shown to be a generic feature of off-shell representations in worldline supersymmetry two decades later. It is shown herein that in all supersymmetric models with spacetime of four or more dimensions, this off-shell supermultiplet twisting, if non-trivial, necessarily maps regular (non-ghost) supermultiplets to ghost supermultiplets. This feature is shown to be ubiquitous in all fully off-shell supersymmetric models with (BV/BRST-treated) constraints.
}
\end{center}
 \vspace{2mm}
\noindent{\footnotesize PACS: 11.30.Pb, 12.60.Jv\hfill
 \parbox[t]{80mm}{\baselineskip=9pt\raggedleft\sl
               Most every ghost will subsist in the twist\,\\*
               of a thought once distraught, but dismissed.\\*[1pt]
               |~Algernon Eduard Beaugh\,}}

\section{Introduction, Results and Conventions}
\label{s:IRS}
The {\em\/twisted\/} variant of well-known chiral off-shell supermultiplets and superfields in 2-dimensional $(2,2)$-supersymmetric field theories were discovered in 1984\cite{rTwSJG0}. Ref.\cite{rGHR} then proved that chiral and twisted chiral superfields jointly provide the singularly exceptional means for constructing worldsheet models with non-K\"ahler target spaces.

Refs.\cite{rHSS,r6-1} show that this {\em\/twisting\/} amounts to changing the sign of a single real (Hermitian) supercharge component. The resulting supermultiplet is not equivalent to the original precisely when
 ({\small\bf1})~the number $N$ of real supercharge components is divisible by four and
 ({\small\bf2})~the twisting changes the sign of a real supercharge involved in an odd number of $\ZZ_2$-projections used to obtain the given supermultiplet from an {\em\/intact\/}\ft{Herein, the adjective ``{\em\/intact\/}'' will stand for ``unprojected, unconstrained, ungauged,'' so intact supermultiplets are the same as defined by the original Salam-Strathdee superfields\cite{rSSSS4,r1001,rBK}, with no additional (reality, chirality, bisection, gauge, equivariance, \etc) condition imposed.} one\cite{r6-3.1}. Besides the single familiar case of (twisted) chiral worldsheet supermultiplets\cite{rTwSJG0,rGHR}, this applies to a large fraction of the ${\sim}\,10^{12}$ types of supermultiplet chromotopology for $N\,{\leqslant}\,32$\cite{r6-3.1}.

In all spacetimes of dimension $d\geqslant4$, the number of components in all spinors|and so also the supercharges|is divisible by four.
 Nevertheless, supermultiplet {\em\/twisting\/}, which the above-cited works prove must exist, has never been observed in higher-dimensional supersymmetric models. It is thus natural to ask, ``Where have all the twisted supermultiplets gone?''

\paragraph{Results:} The question is about off-shell supermultiplets, which are well understood only when the number of supercharges is low and/or additional symmetries heavily restrict the model\ft{Off-shell supermultiplet classification is far from complete\cite[\,\,and references therein]{r6-3.1,rKT07}---unlike the representation theory of Lie algebra\cite{rWyb,rHall}, superlagebras\cite{rSch-SAlg,rFSS-SADict} and of on-shell supermultiplets\cite{rYM97Gau,rDF,rVSV}. Standard representation theory organizes eigenspaces of mutually commuting (even) algebra elements. However, off-shell representations must not be characterized as eigenstates of the Hamiltonian, although the Hamiltonian is a central element of the simplest type of supersymmetry algebras\eq{e:SuSyQ}; any such eigenspace statement would constitute a spacetime differential equation, derivable as an Euler-Lagrange equation, and so make the eigenspace an on-shell representation.}. Indeed, no off-shell completion is known for many of the desired supersymmetric models in higher dimensional spacetime. In turn, most of these theories involve various kinds of gauge symmetry and other constraints, for which the BRST and BV formalisms\cite{rBRS,rT-BRS,rBV-Q2,rSW2,rWS-Fields,rMS-QFT,rEZ-QFT3} introduce various corresponding ghost fields\ft{All commuting/bosonic (anticommuting/fermionic) fields spanning Lorentz tensors (spinors) are herein called {\em\/regular\/}; by {\em\/ghosts\/}, I shall indiscriminately mean all fields with the opposite (``wrong'') spin-statistics correspondence.}, all of which then must form off-shell ghost supermultiplets.
The two topics turn out to be related, and Section~\ref{s:Twist} herein proves:
\begin{thrm}[Spin-Statistics of Twisting]\label{T:1}
Every adinkraic\ft{A worldline supermultiplet is ``{\em\/adinkraic\/}'' if it admits a basis of component fields such that every supercharge component turns every component field into precisely one other component field or its derivative\cite{r6-1}. By extension, a higher-dimensional supermultiplet is adinkraic if its worldline dimensional reduction is.} off-shell supermultiplet of $N$-com\-po\-nent supersymmetry with no central charge in $d$-dimensional spacetime has a twisted variant.
 If not equivalent to the original by field redefinitions, the twisted variant consists solely of ghost component fields for $d\,{\geqslant}\,4$.
Conversely, each off-shell ghost adinkraic supermultiplet in $d\geqslant4$ spacetime is the twisted variant of a regular such supermultiplet.
\end{thrm}
The Lorentz symmetry $\Spin(1,d{-}1)$ is abelian for $d\leqslant2$, and component fields may be chosen to be either regular or ghost fields regardless of their spin (Lorentz-boost) eigenvalue.
Spacetimes of $d\,{=}\,3$ dimension are exceptional in that they admit {\em\/anyons\/}, and will not be discussed herein.

As defined in Refs.\cite{rTwSJG0,rGHR,rHSS,r6-1}, supermultiplet twisting cannot be performed in $(d\,{\geq}\,4)$-di\-men\-si\-onal spacetimes where Lorentz-covariant supercharges have an even number of real components, so a single real component cannot flip sign in a Lorentz-covariant way. For Theorem~\ref{T:1},
Section~\ref{s:Twist} then provides a general Definition~\ref{D:Tw} of supermultiplet twisting, the unique one that:
 ({\small\bf1})~applies to all supersymmetric models in all $d$-dimensional spacetimes, and
 ({\small\bf2})~agrees with the original definition\cite{rTwSJG0,rGHR,rHSS,r6-1} when dimensionally reduced to $(d\,{\leq}\,2)$-dimensional spacetime.

 Section~\ref{s:X10d} shows that Theorem~\ref{T:1} in fact applies much more generally, to all representations that can be constructed by linear algebra from adinkraic representations as building blocks:
\begin{corl}\label{C:1}
Extending by linearity, the statement of Theorem~\ref{T:1} applies to all supermultiplets that may be constructed from adinkraic supermultiplets by linear algebra, that is, either by using superderivative constraints and equivalence classes as in the semi-indefinite sequences of Ref.\cite{r6-1,rTHGK12}, by limited adaptations of Weyl's iterative construction of representations\cite{rH-WWS,rGHHS-CLS}, or by continuous ``entangling''\cite{rGIKT12}.
\end{corl}
The statement of Theorem~\ref{T:1} may well in fact be universal.
 
Section~\ref{s:G} discusses the above results and their relevance within the familiar context of supersymmetric Yang-Mills theories in 4-dimensional spacetime. Section~\ref{s:C} then summarizes the above main results, provides further justification for them, and collects a few concluding remarks. A few necessary technical details are deferred to the appendix.

\paragraph{Conventions:}
The worldline $N$-extended supersymmetry algebra with no central charge is
\begin{equation}
 \left.
 \begin{aligned}
 \big\{\, Q_I \,,\, Q_J \,\big\}&=2\d_{IJ}\,H,&
 \big[\, H \,,\, Q_I \,\big] &=0,\\
 Q_I^{~\dagger}&= Q_I,& H^\dag&=H,\quad
\end{aligned}\right\}\quad
 I,J=1,2,\cdots,N,
  \label{e:SuSyQ}
\end{equation}
were $H=i\vdt$.
 Although not necessary, superspace methods incur no loss of generality\cite{rHTSSp08} and also involve the superderivatives:
\begin{equation}
 \left.
 \begin{aligned}
 \big\{\, \rD_I \,,\, \rD_J \,\big\}&=2i\d_{IJ}\,\vdt,&
 \big[\, \vdt \,,\, \rD_I \,\big] &=0,\\
 \rD_I^{~\dagger}&=-\rD_I,&
 \big\{\, Q_I \,,\, \rD_J \,\big\}&=0,\quad
\end{aligned}\right\}\quad
 I,J=1,2,\cdots,N,
  \label{e:SuSyD}
\end{equation}
such that
\begin{equation}
  Q_I=i\rD_I+2i\d_{IJ}\q^J\,\vdt,\quad\text{and}\quad
  \rD_I=-iQ_I+2i\d_{IJ}\q^J\,H,
 \label{e:Q=iD}
\end{equation}
where the fermionic coordinates $\q^I$ extend (space)time into superspace. When acting on superfields (functions over superspace), the $\rD_I$ act as left-derivatives while the $Q_I$ act as right-derivatives. Therefore, the relationship\eq{e:Q=iD} implies:
\begin{equation}
   Q_I(b) \Defl  i\rD_I\,\bs{\mathcal{B}}| \qquad\text{and}\qquad
   Q_I(f) \Defl -i\rD_I\,\bs{\mathcal{F}}|,
 \label{e:Q=iDBF}
\end{equation}
where $b\Defl\bs{\mathcal{B}}|$ is an arbitrary bosonic functional-differential expression, $f\Defl\bs{\mathcal{F}}|$ a fermionic one, and $\bs{\mathcal{B}}$ and $\bs{\mathcal{F}}$ are the appropriate superderivative superfunctional expressions that define $b$ and $f$, respectively, by setting $\q^I\to0$ as denoted by the right-delimiting ``$|$.''

\section{Supermultiplet Twisting}
\label{s:Twist}
The original example of supermultiplet twisting in Refs.\cite{rTwSJG0,rGHR} pertains to the chiral and twisted chiral superfield pair. Rewriting the usual, 2-component complex Weyl spinor notation $\bDb_\pm\F=0$ and $\bDb_+\X=0=\bD_-\X$ in a suitable real basis, these are defined as:
\begin{alignat}9
  \Rx{\bsf chiral:}\qquad
  [\rD_1-i\rD_3]\F&=&0&=[\rD_2-i\rD_4]\F, \label{e:C}\\*
  \Rx{\bsf twisted chiral:}\qquad
  [\rD_1-i\rD_3]\X&=&0&=[\rD_2+i\rD_4]\X. \label{e:tC}
\end{alignat}
The {\em\/supermultiplet twisting\/} $\F\iff\X$ is thus equivalent to the sign-change
\begin{equation}
  \rD_4\to-\rD_4,\quad\text{whereby also}\quad Q_4\to-Q_4.
 \label{e:tSCh}
\end{equation}
This cannot be compensated by component field redefinitions alone\cite{r6-1}.

Sign-changes such as\eq{e:tSCh} are possible in the worldsheet and worldline (space)times of $d\leqslant2$ dimensions. For $d>2$, minimal spinor representations of the Lorentz group $\Spin(1,d{-}1)$ have an even number of real components, and an odd number of supercharge components cannot change sign in a Lorentz-covariant way.

\paragraph{Strategy:}
We thus seek an off-shell supermultiplet twisting
 equivalent to swapping\eq{e:C}$\,{\iff}\,$(\ref{e:tC})
 but which is Lorentz-covariant for all $d\geqslant0$.
To that end, we use that\eqs{e:C}{e:tC} are adinkraic, identify a general combinatorial criterion (Definition~\ref{D:CP}) for twisting, and prove that another operation (Definition~\ref{D:Tw}, unrestricted by spacetime dimension) implements twisting both by this general combinatorial criterion as well as by dimensional reduction to\eq{e:tSCh}.

\paragraph{Adinkras:}
A judicious definition of real component fields and the conventions of Refs.\cite{r6-1,r6-3.1} allow depicting the supermultiplets\eqs{e:C}{e:tC} faithfully by the Adinkras:
\begin{equation}
 \vC{\begin{picture}(55,36)
   \put(0,0){\includegraphics[width=50mm]{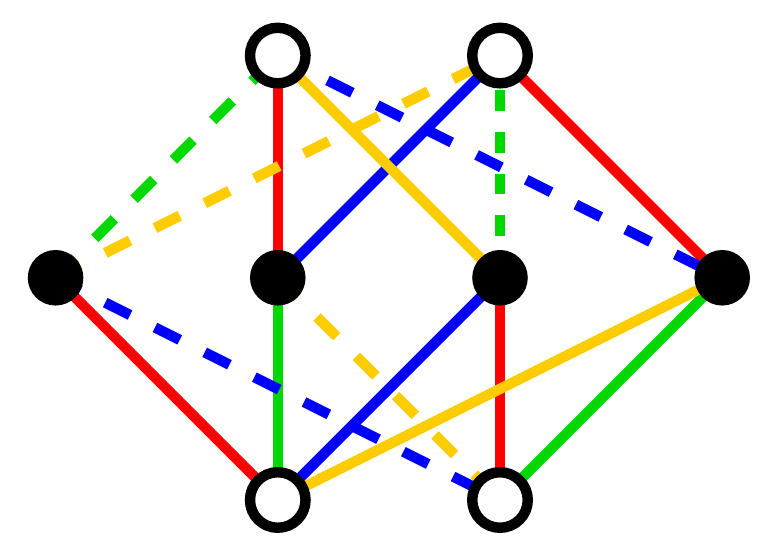}}
     \put(0,28){\Rx{\large$\F$:}}
     \put(15,32){\Rx{$F_3$}}
     \put(35,32){\Lx{$F_4$}}
     \put(1.5,17){\Rx{$\j_1$}}
     \put(15.5,17){\Rx{$\j_2$}}
     \put(34.5,17){\Lx{$\j_3$}}
     \put(48.5,17){\Lx{$\j_4$}}
     \put(15,2){\Rx{$\f_1$}}
     \put(35,2){\Lx{$\f_2$}}
 \end{picture}}
  \qquad\textit{vs.}\qquad
 \vC{\begin{picture}(55,36)(-5,0)
   \put(0,0){\includegraphics[width=50mm]{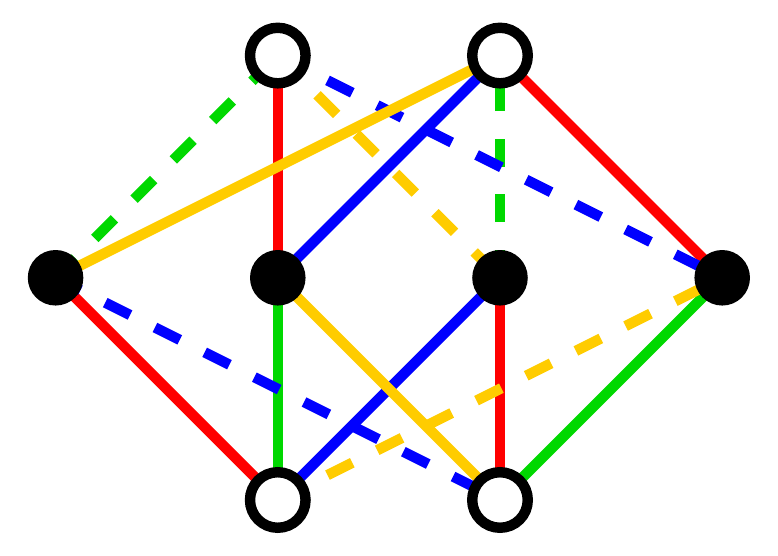}}
     \put(0,28){\Rx{\large$\X$:}}
     \put(15,32){\Rx{$X_3$}}
     \put(35,32){\Lx{$X_4$}}
     \put(1.5,17){\Rx{$\x_1$}}
     \put(15.5,17){\Rx{$\x_2$}}
     \put(34.5,17){\Lx{$\x_3$}}
     \put(48.5,17){\Lx{$\x_4$}}
     \put(15,2){\Rx{$x_1$}}
     \put(35,2){\Lx{$x_2$}}
 \end{picture}}
 \label{e:CtCA}
\end{equation}
The white (black) nodes in an Adinkra depict bosons (fermions); edges of a particular color depict supersymmetry transformations by a corresponding supercharge (emanating clockwise from the $\f_1$ and $x_1$ nodes): \C3{red\,=\,$Q_1$}, \C1{green\,=\,$Q_2$}, \C6{blue\,=\,$Q_3$}, \C8{yellow\,=\,$Q_4$}; Table~\ref{t:A} provides a precise dictionary pertaining to the dimensional reduction of supermultiplets to their worldline.
\begin{table}[ht]
  \centering
  \begin{tabular}{@{} cc|cc @{}}
    \makebox[15mm]{\bsf Adinkra} & \makebox[40mm]{\bsf{\slshape Q}\,-action} 
  & \makebox[15mm]{\bsf Adinkra} & \makebox[40mm]{\bsf{\slshape Q}\,-action} \\ 
    \toprule
    \begin{picture}(5,9)(0,5)
     \put(0,0){\includegraphics[height=11mm]{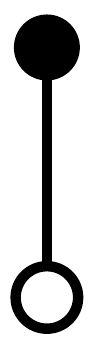}}
     \put(3,0){\scriptsize$A$}
     \put(3,9){\scriptsize$B$}
     \put(-1,4){\scriptsize$I$}
    \end{picture}\vrule depth4mm width0mm
     & $Q_I\begin{bmatrix}\j_B\\\f_A\end{bmatrix}
           =\begin{bmatrix}i\dt\f_A\\\j_B\end{bmatrix}$
  & \begin{picture}(5,9)(0,5)
     \put(0,0){\includegraphics[height=11mm]{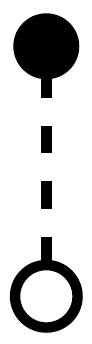}}
     \put(3,0){\scriptsize$A$}
     \put(3,9){\scriptsize$B$}
     \put(-1,4){\scriptsize$I$}
    \end{picture}\vrule depth4mm width0mm
     & $Q_I\begin{bmatrix}\j_B\\\f_A\end{bmatrix}
           =\begin{bmatrix}-i\dt\f_A\\-\j_B\end{bmatrix}$ \\[4mm]
    \midrule
    \begin{picture}(5,9)(0,5)
     \put(0,0){\includegraphics[height=11mm]{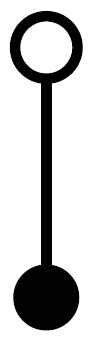}}
     \put(3,0){\scriptsize$B$}
     \put(3,9){\scriptsize$A$}
     \put(-1,4){\scriptsize$I$}
    \end{picture}\vrule depth4mm width0mm
     &  $Q_I\begin{bmatrix}\f_A\\\j_B\end{bmatrix}
           =\begin{bmatrix}\dt\j_B\\i\f_A\end{bmatrix}$
  & \begin{picture}(5,9)(0,5)
     \put(0,0){\includegraphics[height=11mm]{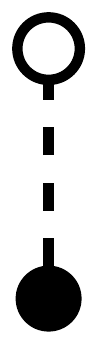}}
     \put(3,0){\scriptsize$B$}
     \put(3,9){\scriptsize$A$}
     \put(-1,4){\scriptsize$I$}
    \end{picture}\vrule depth4mm width0mm
     &  $Q_I\begin{bmatrix}\f_A\\\j_B\end{bmatrix}
           =\begin{bmatrix}-\dt\j_B\\-i\f_A\end{bmatrix}$ \\[4mm]
    \bottomrule
  \multicolumn{4}{c}{\vrule height2.0ex width0pt\small Edges may be drawn in the $I^{\text{th}}$ color instead of being labeled by $I$.}
  \end{tabular}
  \caption{Adinkras assign:
  (white/black) vertices to (boson/fermion) component fields;
  $I^\text{th}$ edge color to $Q_I$;
  solid/dashed edge to $\pm1$ signs in the tabulated supersymmetry action;
  nodes are drawn at heights equal to the mass-dimension of the depicted component (super)field.}
  \label{t:A}
\end{table}
The $\rD_4\to-\rD_4$ ($Q_4\to-Q_4$) twisting is then evident on comparing the two Adinkras\eq{e:CtCA}: they differ solely in the dashedness of the yellow edges depicting the $Q_4$-transformations, and so the sign-differences in table~\ref{t:A}.

The same Adinkra may also be used to depict the supersymmetry transformations in higher-dimensional spacetimes|provided the transformations depicted by the Adinkra are consistent with the additional generators of the higher-dimensional Poincar\'e group, $\Spin(1,d{-}1)\ltimes\IR^{1,d{-}1}_{\,\sss\text{transl.}}$. Conditions for this consistency and therefore dimensional extension to supersymmetry in higher-dimensional spacetimes are being actively investigated\cite{rH-WWS,rFIL,rFL,rGH-obs,rJP-DBt}.

\paragraph{Cycle Parity:}
Besides depicting the system of $(4{+}4){\times}4=32$ explicit component-wise supersymmetry transformation equations in an intuitive but precise 1--1 way, Adinkras permit graph-theoretic and combinatorial methods, which apply to arbitrarily large Adinkras and for arbitrary number of supercharges. Formalizing some observations made in Refs.\cite{r6-3.1,r6-1.2}, we have:
\begin{defn}\label{D:CPC}
Let \bc\ denote an ordered collection of distinct edge-colors. 
 Within an Adinkra, the {\em\/cycle parity\/} of any cycle (closed path) $\sC_\bc$ of \bc-colored edges (one edge for every color in \bc) is the product
\begin{equation}
  \CP(\sC_\bc) \Defl
  (-1)^{\ve_*}\cdot(-1)^{\ve_\bc(\sC_\bc)}\cdot
  \prod_{\text{edge}\,\in\,\sC_\bc}(-1)^{\ve(\text{edge})}
 \label{e:CP}
\end{equation}
 where
 $\ve_*=$\,0~(\,$1$) if\/ $\sC_\bc$ starts from a bosonic (fermionic) node\ft{The $(-1)^{\ve_*}$ factor in the definition of $\CP(\sC_\bc)$\ is the same relative sign between the two expressions in Eqs.\eq{e:Q=iDBF}.},
 $\ve_\bc(\sC_\bc)=0$~(\,$1$) if\/ $\sC_\bc$ follows an even (odd) permutation of \bc,
 and $\ve(\text{edge})=$\,0~(\,$1$) if the edge is solid (dashed).
\end{defn}
That is, given any closed path $\sC$ of distinct edges (supersymmetry transformations), pick some ordering \bc\ of the supersymmetries acting along $\sC$. Start with a factor $+1$ ($-1$) if starting from a bosonic (fermionic) node, multiply by $-1$ if the ordering of the supersymmetry transformations along $\sC$ is an odd permutation of the \bc-ordering, and a $-1$ for every dashed edge in the path; see table~\ref{t:A}. The resulting product is $\CP(\sC_\bc)$.

The Adinkras\eq{e:CtCA} make it easy to compute (pick \bc\,=\,\{\C3{1}, \C1{2}, \C6{3}, \C8{4}\}, say)
\begin{equation}
  \CP(\sC_\bc)=+1,~~\forall\>\sC_\bc\subset\fA(\F),
   \qquad\text{and}\qquad
  \CP(\sC_\bc)=-1,~~\forall\>\sC_\bc\subset\fA(\X).
 \label{e:CP4}
\end{equation}
We now prove that $\CP(\sC_\bc)$ is a characteristic of any part of the Adinkra $\fA$ reachable by the $Q$-transformations used in $\sC_\bc$, and so is independent of the particular cycle $\sC_\bc$ and starting node used to compute it.

\begin{proof}
The values\eq{e:CP4} may in fact be computed directly from the superspace definitions\eqs{e:C}{e:tC}, as follows: Denoting the real and imaginary parts of the chiral superfield as $\F=\IA+i\mkern2mu\IB$, the complex system of superderivative constraints\eq{e:C} is seen to be equivalent to the real system
\begin{subequations}
 \label{e:CDs}
\begin{alignat}9
 \big(\C3{\rD_1}\IA+\C6{\rD_3}\IB\big)&=0&
 &=i\big(\C3{\rD_1}\IB-\C6{\rD_3}\IA\big),\\
 \big(\C1{\rD_2}\IA+\C8{\rD_4}\IB\big)&=0&
 &=i\big(\C1{\rD_2}\IB-\C8{\rD_4}\IA\big).
\end{alignat}
\end{subequations}
Elimination of $\IB$, and then in turn of $\IA$ (see the appendix for details) imply, respectively,
\begin{subequations}
% \label{e:}
\begin{equation}
  \ID^+_{IJ}\,\IA=0,
   \quad\text{and}\quad
  \ID^+_{IJ}\,\IB=0,
 \label{e:SD4+}
\end{equation}
with
\begin{equation}
   \ID^\pm_{IJ}\Defl[\rD_I\rD_J\pm\fc12\ve_{IJ}{}^{KL}\rD_K\rD_L],
 \label{e:DIJ}
\end{equation}
so that the projection constraint\eq{e:SD4+} equally applies to $\F=(\IA+i\IB)$. A second application of $\ID^+_{IJ}$ on $\F$ then produces:
\begin{align}
  \big[\vdt^2+\C3{\rD_1}\C1{\rD_2}\C6{\rD_3}\C8{\rD_4}\big]\F&=0,
   \label{e:dD4+}\\*
   \qquad\overset{(\ref{e:Q=iDBF})}{\Longrightarrow}\qquad
  \C3{Q_1}\,\C1{Q_2}{\circ}\C6{Q_3}{\circ}\C8{Q_4}{\circ}\F&=+(H^2\,\F).
  \qquad\qquad \label{e:HQ4+}
\end{align}
\end{subequations}
The analogous computation for $\X$ yields:
\begin{subequations}
% \label{e:}
\begin{align}
  \ID^-_{IJ}\,\X&=0, \label{e:SD4-}\\
  \To\quad\big[\vdt^2-\C3{\rD_1}\C1{\rD_2}\C6{\rD_3}\C8{\rD_4}\big]\X&=0,
   \label{e:dD4-}\\*
   \qquad\overset{(\ref{e:Q=iDBF})}{\Longrightarrow}\qquad
  \C3{Q_1}{\circ}\C1{Q_2}{\circ}\C6{Q_3}{\circ}\C8{Q_4}\,\X&=-(H^2\,\X).
   \qquad\qquad
 \label{e:HQ4-}
\end{align}
\end{subequations}
We thus conclude:
\begin{enumerate}\itemsep=-3pt\vspace{-1mm}
 \item The 4-color path depicting the $\C3{Q_1}{\circ}\C1{Q_2}{\circ}\C6{Q_3}{\circ}\C8{Q_4}$ action in the Adinkra is closed precisely if the supermultiplet satisfies a superderivative system of constraints akin to\eq{e:SD4+} and\eq{e:SD4-}.
 \item The relative signs in the superderivative binomial operators\eq{e:SD4+} and\eq{e:dD4+} equal the sign on the right-hand side of\eq{e:HQ4+} and the sign computed for $\F$ by Definition~\ref{D:CPC}; the corresponding signs in\eq{e:SD4-}, \eq{e:dD4-} and\eq{e:HQ4-} equal  the sign for $\X$ by the same definition.
\end{enumerate}

The lowest components of both $\F$ and $\X$ are complex bosons. The superspace\,$\to$\,(space)time projection of\eq{e:HQ4+} and\eq{e:HQ4-} immediately reads off the value of $\CP(\sC_\bc)=+1$, \ie, $\CP(\sC_\bc)=-1$, where $\sC_\bc$ is the 4-cycle of edges starting and ending at either the real or the imaginary part of the lowest bosonic component field, and colored in a permutation of \bc\,=\,\{\C3{red}, \C1{green}, \C6{blue} and \C8{yellow}\}.
 Since distinct $\rD_I$'s and $Q_I$'s all anticommute, the $\rD_I$- and the $Q_I$-monomials in\eq{e:dD4+}, \eq{e:HQ4+}, \eq{e:dD4-} and\eq{e:HQ4-} may be freely permuted, keeping track of the resulting change in the relative sign by $(-1)^{\ve_\bc(\sC)}$.
 Finally, all other component fields are computed by projecting a $\rD_{\sss[I_1}\cdots\rD_{\sss I_n]}$-superderivative of $\F$, \ie, $\X$. Since $I_1,\cdots,I_n\in\{\C31,\C12,\C63,\C84\}$,
\begin{equation}
   \rD_{\sss[I_1}\cdots\rD_{\sss I_n]}\big[H^2\pm
                                  \C3{\rD_1}\C1{\rD_2}\C6{\rD_3}\C8{\rD_4}\big]
   =\big[H^2\pm
         (-1)^n\C3{\rD_1}\C1{\rD_2}\C6{\rD_3}\C8{\rD_4}\big]
          \rD_{\sss[I_1}\cdots\rD_{\sss I_n]}.
 \label{e:DD4}
\end{equation}
The sign computed starting from a fermionic component field (odd $n$) is then opposite of the one computed by starting from a bosonic component field (even $n$).

 Generalizations to arbitrary cycles $\sC_\bc$ in other Adinkras replace the quartic $\rD_I$- and $Q_I$-monomials from the computations\eqs{e:SD4+}{e:DD4} with the $\sC_\bc$-depicted $\rD_I$- and $Q_I$-monomials and their permutations---and\eq{e:DD4} is restricted to component fields that can be reached one from another using only the $\sC_\bc$-depicted $Q_I$'s.
 Every closed path $\sC_\bc$ of non-repeating edge-colors in any such sub-Adinkra thus defines a $\bc$-ordered superderivative $\rD_{\sss[I_1},\cdots,\rD_{\sss I_n]}$ for $I\in\bc$, and thus a corresponding system of self-(anti)dual operators generalizing\eq{e:DIJ}. In turn, these self-(anti)dual operators define superfield representations\cite{r6-1.2} for all $\sim10^{12}$ chromotopology types of adinkraic supermultiplets\cite{r6-3.1}.

Within any of these sub-Adinkras, the analogue of the above computation thus computes $\CP(\sC_\bc)$ for each part connected by $\{Q_I:I\in\bc\}$ and is independent of the particular cycle $\sC_\bc$ within this part as well as the starting node used to compute it.
\end{proof}

\begin{defn}\label{D:CP} The cycle parity
$\CP_\bc(\fA)=\frac1{\#(\sC_\bc)}\sum_{\sC_\bc\subset\fA}\CP(\sC_\bc)$ is a characteristic of the supermultiplet depicted by the Adinkra \fA; if \fA\ contains no \bc-colored cycle, $\CP_\bc(\fA)\Defl0$.
\end{defn}
By the so-called Burnside's Lemma\cite{rPC-Cx,rRS-Cx,rR+T-Combi}, the computation of the sum in this definition may be partitioned into disjoint sub-Adinkras of the Adinkra \fA\ that are connected by \bc-edges only. The preceding proof guarantees that any one cycle within each disjoint sub-Adinkras suffice for computing $\CP(\sC_\bc)$ within each such sub-Adinkra, so that $\#(\sC_\bc)$ reduces to the number of such sub-Adinkras.
 That is, we simply delete all other edges, compute $\CP(\sC_\bc)$ within each disjoint sub-Adinkra using any one representative cycle $\sC_\bc$, and average over the so disjoined sub-Adinkras.

Similar quantities have been defined elsewhere in the literature: For any adinkraic supermultiplet,
 the combinatorially defined $\CP_\bc(\fA)$ coincides with
 the algebraic characteristic $\c_0$ defined for $N\,{=}\,4$ in Ref.\cite{rUMD09-1}, and used there to detect off-shell supermultiplet twisting. In turn, $\CP_\bc(\fA)$ is defined for arbitrary $N$ and is also well-defined for non-adinkraic representations such as the supermultiplets discussed in Refs.\cite{rTHGK12,rGHHS-CLS,r6-4.2}.

\paragraph{Generality:}
Systems of superderivative constraints akin to\eq{e:SD4+} were used in Ref.\cite{r6-1.2} to construct a super-constrained superfield array for each of ${\sim}\,10^{12}$ chromotopologies of Ref.\cite{r6-3.1}, whereupon Theorem~7.6 of Ref.\cite{r6-1} constructs a superfield representation for every supermultiplet with that chromotopology. The particular superderivative $\rD_{\sss[I_1}\cdots\rD_{\sss I_n]}$ occurring in the induced superderivative constraint analogue of\eq{e:dD4+} specifies the chromotopology of the so-defined superfield and supermultiplet. For the present purposes a comparison of Eqs.\eq{e:dD4+} and\eq{e:dD4-} makes it obvious that $\F$ and $\X$ satisfy complementary projections.

The classification work of Ref.\cite[ and references therein]{r6-3.1} and the explicit construction in Ref.\cite{r6-1.2} imply for each $N$ that there is a single intact ($N$-cubical) chromotopology, but a combinatorially growing number of (multiple) $\ZZ_2$-projections, where each projection is encoded by superderivative constraints akin to \Eqs{e:SD4+}{e:dD4+} and \Eqs{e:SD4-}{e:dD4-}.
 Supermultiplets with a multiple $\ZZ_2$-projection chromotopology will satisfy {\em\/multiple\/} independent projections akin to that in\eq{e:dD4+} and\eq{e:dD4-}. Flipping the relative sign in {\em\/two\/} independent such projections is equivalent to changing the sign of {\em\/two\/} of the supercharges, which can always be compensated by judicious sign-changes in some of the component fields.
 
It then follows that every Adinkra with a chromotopology of a (multiple) $\ZZ_2$-projection of an $N$-cube has at most one pair of mutually twisted variants. In turn, the twisting in adinkraic supermultiplets cannot be compensated by a judicious component field redefinition precisely when $N\equiv0\,\text{(mod~4)}$\cite{r6-3.1}.
 The original and the twisted supermultiplets are inequivalent precisely if the twisting changes the sign of an odd number of supercharges involved in an odd number of distinct projections.
 The vast majority of the ${\sim}\,10^{12}$ chromotopologies are projected and so do have inequivalent twisted variants.

Thus, all adinkraic worldline dimensional reductions of supermultiplets in spacetimes of dimension four and higher must have a twisted variant, most of which inequivalent to the original.
 In all worldline and all worldsheet theories, the sign-change twisting akin to\eq{e:tSCh} generates all twisted supermultiplets.
 However, this operation:
\begin{enumerate}\itemsep=-3pt\vspace{-2mm}
 \item is global in the sense of Ref\cite{rGIKT12}: it affects all supermultiplets at once,
 \item would violate Lorentz covariance in spacetimes of $d\geqslant2$ dimension.
\end{enumerate}
We now find an equivalent twisting which turns out to avoid both of these problems.

\paragraph{Equivalent Twisting:}
Since the sign-change of $\CP_\bc(\fA)$ detects the twisting of the off-shell supermultiplet depicted by the Adinkra \fA, the factor $(-1)^{\ve_*}$ in the Definition~\ref{D:CPC} implies that the twisted variant of any given Adinkra $\fA$ is equivalently obtained by swapping the spin-statistics (boson\,$\iff$\,fermion, \ie, commuting\,$\iff$\,anticommuting) of all nodes, sometimes referred to as the ``Klein-flip,'' $\fA^{\sss K}$. That is to say,
\begin{equation}
 \CP_\bc(\fA^{\sss K})=-\CP_\bc(\fA), \label{e:KFA}
\end{equation}
and supermultiplet twisting is {\em\/equivalent\/} to Klein-flipping the supermultiplet, up to some component field redefinitions.

For the particular case\eqs{e:C}{e:tC}, this is verified explicitly by identifying:
\begin{equation}
 \X\too{\sss K}\F:\quad\bigg\{
 \begin{array}{r@{~\iff~}lr@{~\iff~}lr@{~\iff~}lr@{~\iff~}l}
  \f_1&\x_2^{\sss K},&\quad
  \f_2&\x_4^{\sss K},&\quad
   F_3&\dt\x_1^{\sss K},&\quad
   F_4&\dt\x_3^{\sss K},\\*[2mm]
  \j_1& X_3^{\sss K},&\quad
  \j_2&\dt{x}_1^{\sss K},&\quad
  \j_3& X_4^{\sss K},&\quad
  \j_4&\dt{x}_2^{\sss K}.
 \end{array}
 \label{e:X>F}
\end{equation}
Owing to the non-local field redefinitions such as $F_4=\dt\x_3^{\sss K}$ (whereby $\x_3^{\sss K}=\int\!\rd\t\,F_4$), this mapping is not a {\em\/strict homomorphism of off-shell supermultiplets\/}\cite[~Definition~B.1]{r6-3.2}, and $\F$ and $\X$ are not isomorphic as representations of supersymmetry, but are each other's twisted variant. The replacements
  $x_1\to\dt{x}^{\sss K}_1\mapsto\j_2$, 
   $x_2\to\dt{x}^{\sss K}_2\mapsto\j_4$, 
    $\x_1\to\dt\x^{\sss K}_1\mapsto F_3$ and 
     $\x_3\to\dt\x^{\sss K}_3\mapsto F_4$ are easily depicted:
\begin{equation}
 \vC{\begin{picture}(160,32)(0,-1)
   \put(0,0){\includegraphics[width=160mm]{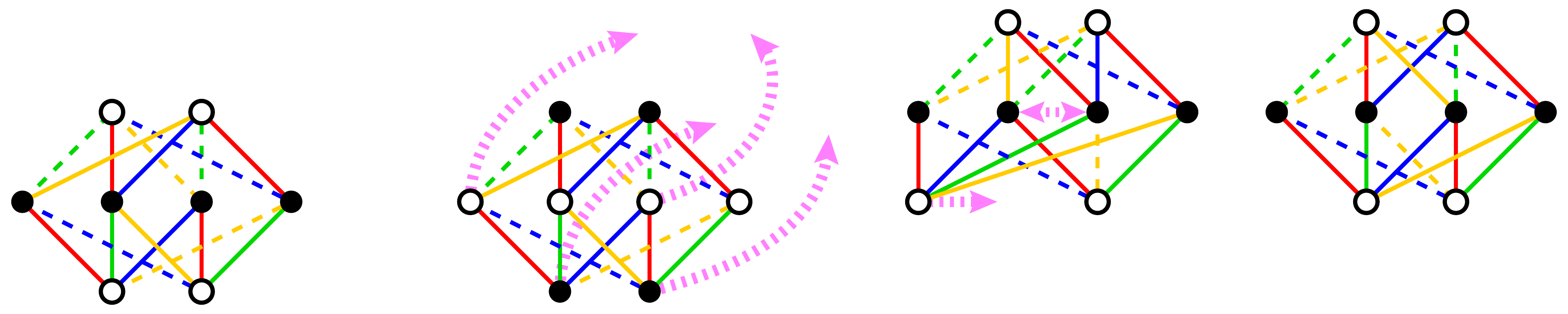}}
     \put(73.5,19){$\SSS\j_2$}
     \put(83,19){$\SSS\j_4$}
     \put(65,28){$\SSS F_3$}
     \put(73.5,28){$\SSS F_4$}
     \put(16,29){\Cx{\small twisted-chiral}}
     \put(16,25){\Cx{\footnotesize right-hand side of\eq{e:CtCA}}}
     \put(38,12.5){\Cx{\scriptsize\sl Klein}}
     \put(33,11.25){\vector(1,0){12}}
     \put(38,8.5){\Cx{\scriptsize\sl flip}}
     \put(144,5){\Cx{\small chiral}}
     \put(144,1){\Cx{\footnotesize left-hand side of\eq{e:CtCA}}}
 \end{picture}}
 \label{e:X>FA}
\end{equation}
This illustrates how the assignments\eq{e:X>F} produce, upon the final Klein flip (boson\,$\iff$\,fermion, \ie, white\,$\iff$\,black node assignment swap), the Adinkra of the chiral supermultiplet, $\F$, in\eq{e:CtCA}.

In fact, by using
\begin{subequations}
 \label{e:SuSyDc}
\begin{equation}
  \bD_+\Defl\C3{\rD_1}+i\C6{\rD_3}, \qquad
  \bD_-\Defl\C1{\rD_2}+i\C8{\rD_4},
 \label{e:SuSyDc1}
\end{equation}
and their conjugates, the algebra\eq{e:SuSyD} extends to the $(2,2)$-supersymmetric worldsheet:
\begin{equation}
  \big\{\,\bD_+\,,\,\bDb_+\,\big\} = \vd_\pp,\quad
  \big\{\,\bD_-\,,\,\bDb_-\,\big\} = \vd_\mm,\quad
  \big\{\,\bD_+\,,\,\bD_-\,\big\} =0= \big\{\,\bD_+\,,\,\bDb_-\,\big\}.
 \label{e:SuSyDc2}
\end{equation}
\end{subequations}
Thus---without redefining the supercharges---the result of\eq{e:X>FA} is equivalent to the original off-shell supermultiplet twisting\cite{rTwSJG0,rGHR,rHSS} even within $d=2$ spacetime such as the worldsheet, where the entire sequence of redefinitions\eq{e:X>FA} is also valid; see appendix~A of Ref.\cite{rH-WWS}.

As only the first step in\eq{e:X>FA} actually changes the sign of $\CP_\bc(\fA)$ and applies in spacetimes of all dimensions (it affects neither Lorentz- nor Poincar\'e-covariance), we adopt:
\begin{defn}[Off-Shell Supermultiplet Twisting]\label{D:Tw}
For any off-shell supermultiplet, its twisted variant is defined to be the result of the boson$\,{\iff}\,$fermion (spin-statistics) Klein-flip.
\end{defn}
The definitions~\ref{D:CPC}--\ref{D:CP} of $\CP_\bc(\fA)$ show no other Lorentz-covariant way to flip the sign of $\CP_\bc(\fA)$  and so extend to $d\geqslant4$ spacetimes the original operation\eq{e:tSCh}, swapping\eqs{e:C}{e:tC}.
 In turn, the redefinitions\eqs{e:X>F}{e:X>FA} and\eq{e:SuSyDc} demonstrate the $d\leqslant2$ equivalence of Definition~\ref{D:Tw} and the original definition of twisting\cite{rTwSJG0,rGHR}---as given and valid only within $d\leqslant2$.
 The foregoing then comprises:

\paragraph{Proof of Theorem~\ref{T:1}:}
The Definition~\ref{D:Tw}:
 ({\small\bf1})~flips the sign of $\CP_\bc(\fA)$, so the worldsheet (worldline) reductions of the original and the resulting supermultiplet are indeed twisted variants of each other;
 ({\small\bf2})~changes the spin-statistics assignment of the component fields but not the Lorentz-representations spanned by them, and so turns a regular off-shell supermultiplet into a ghost one and {\em\/vice versa\/}.
The converse portion of Theorem~\ref{T:1} then must also hold:
 the twisted variant of an off-shell ghost supermultiplet must be a regular off-shell supermultiplet.
\QED

 The twisting in Definition~\ref{D:Tw} swaps boson/fermion, \ie, commuting/anticommuting assignments of all component fields, regardless of the representations of any and all symmetry groups that the fields may span.
 This has been employed in Ref.\cite{rFGH} to construct the twisted variant of the so-called {\em\/ultra-multiplet\/}\cite{rGR0} of 8-component worldline supersymmetry. However, that work considered effective/dynamical symmetries rather than the Lorentz group of spacetime symmetries, and so the twisting did not incur a spin-statistics regular$\,{\iff}\,$ghost flip. Similarly, while the well-known triality of $\Spin(8)$ could circumnavigate the above conclusion, there exists no spacetime in which $\Spin(8)$ is the Lorentz group.

\section{Dimensional Extensions and Other Generalizations}
\label{s:X10d}
The foregoing applies to all adinkraic supermultiplets|in the worldline dimensional reduction of which each supercharge transforms every component field into precisely one other component field or its derivative.

\paragraph{Beyond Adinkras:}
While the simplest and most often used off-shell supermultiplets are indeed adinkraic, they may be used to construct indefinitely many and ever larger off-shell supermultiplets that are not adinkraic\cite{rTHGK12,rGIKT12}, some of which in fact are in current use\cite{rGHHS-CLS,r6-4.2}. We thus use Adinkras as the simpler building blocks from which to construct and analyze indefinitely many other, more complicated off-shell representations.

In particular, Ref.\cite{r6-1} presents a sequence of indefinitely many supermultiplets defined by means of superderivative constraining, generalizing \Eqs{e:C}{e:tC}. In all such constructions, one starts with an array (direct sum) of real, off-shell intact Salam-Strathdee superfields\cite{rSSSS4}, $\IU_a$, and defines a superfield/supermultiplet as the solution of the superderivative system
\begin{equation}
   \IA\,\Defl~\big\{\,\IU_a\>:~\sum\nolimits_a\,\ID_A{}^a\,\IU_a = 0,~~\forall A\,\big\},
 \label{e:DU=0}
\end{equation}
where each $\ID_A{}^a$ is a suitable matrix of formal multinomials in the superderivatives $\rD_I$'s. Formally, the space of solutions of\eq{e:DU=0} is the kernel of the mapping $\ID_A{}^a:\IU_a\to\Tw\IU_A$, the dual of which is the ``gauge'' equivalence class
\begin{equation}
  \IV=\Tw\IU_A/(\ID_A{}^a\,\IU_a)
   \Defl\big\{\,\f_A\in\Tw\IU_A:~\f_A\simeq\f_A+\ID_A{}^a\vf_a,~~\vf_a\in\IU_a)\,\big\}.
 \label{e:U/DU}
\end{equation}
Since $\{Q_I,\rD_J\}=0$, such maps are supersymmetric, so that both $\IA$ and $\IV$ are invariantly defined off-shell supermultiplets.
 Ref.\cite{rTHGK12} verifies that the so-constructed $\IV$'s are most often not adinkraic.
 From the standard theory of linear mappings of algebraic structures, we then have the sequence:
\begin{equation}
  \big\{\,\ker(\ID_A{}^a)=\IA\,\big\} \overset\i\into
   \big\{\,\IU_a\,\big\} \too{~\ID_A{}^a~}
    \big\{\,\IV = \mathop{\textrm{coker}}(\ID_A{}^a)\,\big\}
\end{equation}
where $\i$ is a 1--1 inclusion map and $\ID_A{}^a$ is constructed so that $\ID_A{}^a\circ\i=0$. One also says that $\{\IU_a\}$ is then an {\em\/extension\/} of $\IV$ by $\IA$, which is a non-symmetric notion of a sum.

 Since the $\ID_A{}^a$-map and the constraint equations\eq{e:DU=0} are all linear in $\IU_a$, both $\IA$ and $\IV$ are linearly complementary within $\IU_a$,
 determined by the choice of the $\ID_A{}^a$'s,
 so $\CP_\bc(\IA)$ and $\CP_\bc(\IV)$ may be computed along the lines of\eqs{e:CDs}{e:DD4}.
 It follows that
\begin{equation}
    \CP_\bc(\IA)+\CP_\bc(\IV)=\CP_\bc(\oplus_a\IU_a).
  \label{e:CPA+V}
\end{equation}
Since each $\IU_a$ is an intact supermultiplet, $\CP_\bc(\oplus_a\IU_a)=0$ and $\CP_\bc(\IA)=-\CP_\bc(\IV)$. It is tempting to conjecture that the result\eq{e:CPA+V} generalizes to the cases where $\IU_a$ are replaced by non-intact, distinct and even non-Adinkraic supermultiplets.

\paragraph{A Known Example:}
The result\eq{e:CPA+V} certainly agrees with the simplest construction of this kind, where superderivative constraints generalizing\eq{e:SD4+} and\eq{e:SD4-} were used to reduce the intact supermultiplet $\IU$ to a sub-supermultiplet $\IA_{\sC}$ of any one of the ${\sim}\,10^{12}$ $\sC$-encoded chromotopologies\cite{r6-1.2}. In the simplest non-trivial case, for $N\,{=}\,4$, we have
\begin{gather}
 \vC{\begin{picture}(140,32)(0,3)
      \put(2,2){\includegraphics[height=19mm]{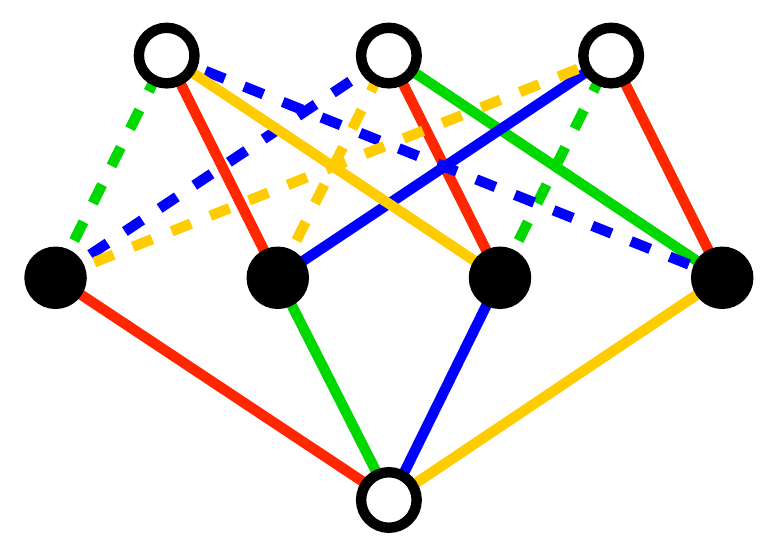}}
       \put(3,4){$\IX$}
       \put(30,18){\Large$\overset\i\into$}
      \put(42,2.1){\includegraphics[height=34mm]{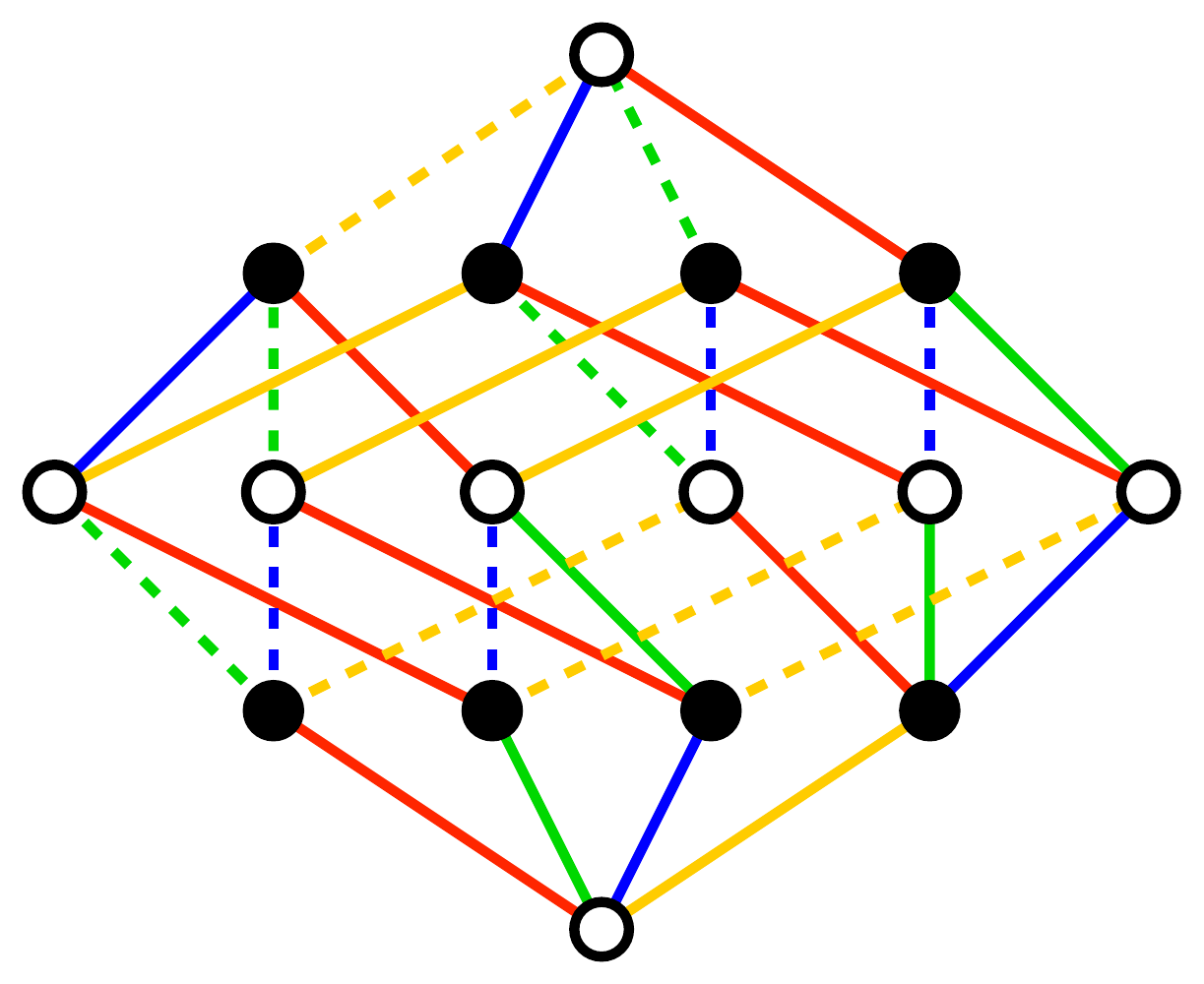}}
       \put(48,4){$\IU$}
       \put(87,18){\Large$\too{~\ID_{IJ}^+~}$}
      \put(104,17.2){\includegraphics[height=19mm]{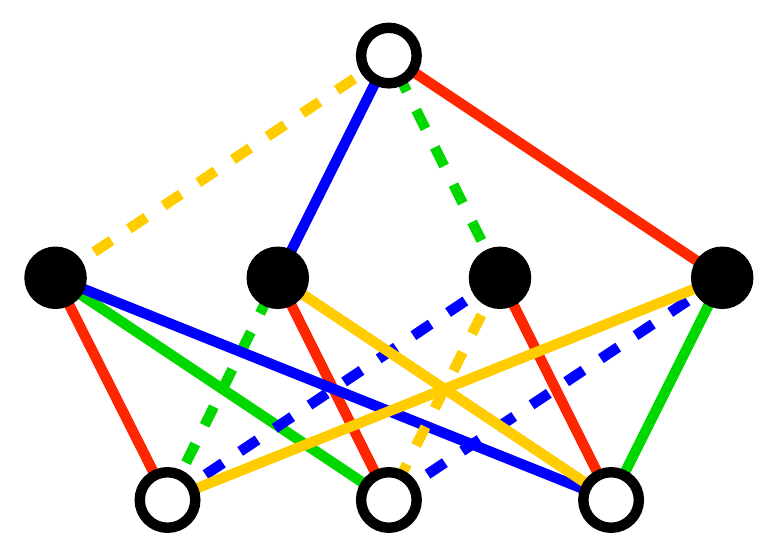}}
       \put(105,30){$\IY$}
     \end{picture}}
 \label{e:XYZ}\\
 \IX \subset \IU:~~ \ID_{IJ}^+\IX=0,
  \qquad\text{and}\qquad
 \IY \Defl \{\IU\simeq\IU+\i(\IX)\},
 \label{e:XYZe}
\end{gather}
using the definition\eq{e:DIJ}. Following a cycle of, say, \bc\,=\,\{\C31, \C12, \C63, \C84\} edges starting from any bosonic (white) node, it is easy to compute $\CP_\bc(\IX)=1$, $\CP_\bc(\IY)=-1$ and $\CP_\bc(\IU)=0$.

As stated above, the structure of the $Q$-action in $\IX$ is determined by $\ID^+_{IJ}$, since
\begin{equation}
 \vC{\begin{picture}(150,30)(-5,-2)
       \put(23.75,0){\includegraphics[height=28mm]{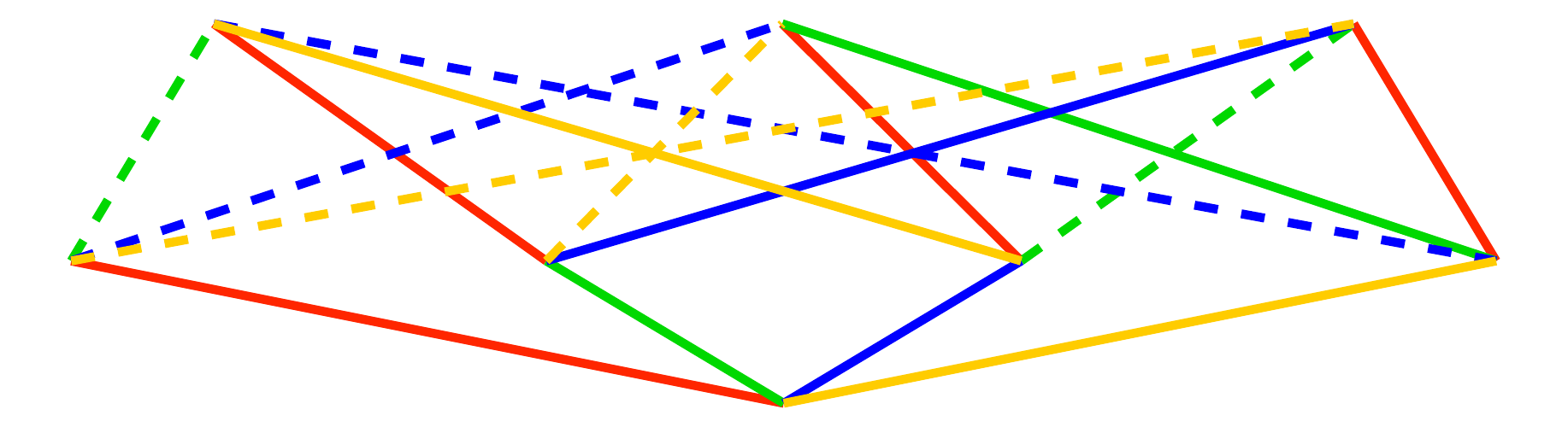}}
        \put(5,25){$\ID^+_{IJ}$:}
        \put(37,25){\cB{$\C3{\rD_1}\C1{\rD_2}=-\C6{\rD_3}\C8{\rD_4}$}}
        \put(75,25){\cB{$\C3{\rD_1}\C6{\rD_3}=+\C1{\rD_2}\C8{\rD_4}$}}
        \put(113,25){\cB{$\C3{\rD_1}\C8{\rD_4}=-\C1{\rD_2}\C6{\rD_3}$}}
        \put(28,10){\cB{\C3{$\rD_1$}}}
        \put(60,10){\cB{\C1{$\rD_2$}}}
        \put(90,10){\cB{\C6{$\rD_3$}}}
        \put(122,10){\cB{\C8{$\rD_4$}}}
        \put(75,0){\cB{$\Ione$}}
     \end{picture}}
 \label{e:DcA}
\end{equation}
The nodes of this {\em\/operatorial\/} Adinkra specifies 1{+}4{+}3 operators, and its edges encode the sign in the successive application of these operators. For example, following the left-most edges from $\Ione$ upward and acting with the $\rD_I$ from the left, we have that
\begin{subequations}
\begin{equation}
  \Ione ~\C3{\underset{\sss\text{solid}}{\too{~+\rD_1~}}}~ {+}\C3{\rD_1}
         ~\C1{\underset{\sss\text{dashed}}{\too{~-\rD_2~}}}~ {-}\C1{\rD_2}\C3{\rD_1}
   = {+}\C3{\rD_1}\C1{\rD_2},
\end{equation}
while following the next to left-most edges one has
\begin{equation}
  \Ione ~\C1{\underset{\sss\text{solid}}{\too{~+\rD_2~}}}~ {+}\C1{\rD_2}
         ~\C3{\underset{\sss\text{solid}}{\too{~+\rD_1~}}}~ {+}\C3{\rD_1}\C1{\rD_2},
\end{equation}
next to that,
\begin{equation}
  \Ione ~\C6{\underset{\sss\text{solid}}{\too{~+\rD_3~}}}~ {+}\C6{\rD_3}
         ~\C8{\underset{\sss\text{solid}}{\too{~+\rD_4~}}}~ {+}\C8{\rD_4}\C6{\rD_3}
   = {-}\C6{\rD_3}\C8{\rD_4}
   \isBy{\sss\text{(\ref{e:SD4+})}} {+}\C3{\rD_1}\C1{\rD_2},
\end{equation}
and so on. It is easy to see that $\CP_{\{\C31,\C12,\C63,\C84\}}(\sC\in\text{(\ref{e:DcA})})=+1$, just as it is for $\IX$. Since $\IU$ is an intact supermultiplet, $\CP_{\sss\{\C31,\C12,\C63,\C84\}}(\IU)=0$, Eq.\eq{e:CPA+V} implies that $\CP_{\sss\{\C31,\C12,\C63,\C84\}}(\IY)=-1$, as is easy to verify. This is the precise sense in which $\CP_\bc(\IX)$ and $\CP_\bc(\IY)$ are both determined by $\ID^+_{IJ}$ and $\CP_\bc(\IU)$.
\end{subequations}

The relationships indicated in\eq{e:XYZ} are in fact fairly well-known in the supersymmetry literature in 1{+}3-dimensional spacetime: The central Adinkra $\IU$ may be identified with the self-conjugate (real), so-called ``vector'' superfield $\IV=\ba\IV$. The Adinkra $\IX$ may be identified with the superfield combination $i(\bs\L{-}\ba{\bs\L})$ where $\bs\L$ is a chiral superfield and $\ba{\bs\L}$ its Hermitian conjugate; see appendix for details.
 This identifies $\IY$ with the gauge equivalence class $\{\IV\simeq\IV{+}i(\bs\L{-}\ba{\bs\L})\}$---the vector superfield in the Wess-Zumino gauge.

Although the component fields of both $\IX$ and $\IY$ obey the standard spin-statistics correspondence, not the ``wrong'' one of ghosts, this does not contradict Theorem~\ref{T:1}, as we now discuss.

\paragraph{Theorem~\ref{T:1} Compliance:}
Although they are related to $\F$ from\eq{e:C} and $\X$ from\eq{e:tC}, respectively, via exclusively worldline-definable non-local field redefinitions such as those in\eq{e:X>F}, the depiction\eq{e:XYZ} clearly shows that $\IX$ and $\IY$ are not each other's image through the twisting operation of Definition~\ref{D:Tw}: their nodes (component fields) have different heights (mass-dimensions).
 
Thus, although the worldline supermultiplets
\begin{equation}
 \begin{gathered}
  \F=\big(\f_1,\f_2\big|\j_I\big|F_3,F_4\big),\quad
  \big(\f_i\big|\j_I\big),\quad
  \IX=\big(\f_1\big|\j_I\big|F_3,\Dt\f_2,F_4\big),\quad\etc\\
  \text{where}\quad \f_3\Defl({\ttt\int\!\rd\t\,F_3}),\quad
   \text{and}\quad \f_4\Defl({\ttt\int\!\rd\t\,F_4}),
 \end{gathered}
 \label{e:cA}
\end{equation}
are all related to each other by means of node-raising/lowering and so have the same supersymmetry transformation structure\ft{Following the formalism of Refs.\cite{r6-3.1}, this structure might be called  {\em\/dashed chromotopology\/}.}, they are:
\begin{enumerate}\itemsep=-3pt%\vspace*{-3mm}
 \item locally inequivalent worldline supermultiplets, and also
 \item worldline reductions of distinct higher-dimensional supermultiplets---if they extend at all to higher-dimensional spacetime\cite{rH-WWS,rFIL,rFL,rGH-obs}.
\end{enumerate}\vspace*{-3mm}
Indeed, this applies just as well to all the sixteen node-raised/lowered variants of\eq{e:cA}:
 starting with $(\f_i|\j_I)$ on the worldline, one can ``raise'' any of the four bosons, obtaining $2^4=16$ Adinkras with different node-height arrangements.
 The analogous applies also to the twisted variants of\eq{e:cA}:
\begin{equation}
 \begin{gathered}
  \X=\big(x_1,x_2\big|\x_I\big|X_3,X_4\big),\quad
  \big(x_i\big|\x_I\big),\quad
  \IY=\big(({-}x_1),x_2,x_3\big|{-}\x_1,\x_2,\x_3,\x_4\big|X_4\big),\\
  \text{where}\quad x_3\Defl({\ttt\int\!\rd\t\,X_3}),\quad
   \text{and}\quad x_4\Defl({\ttt\int\!\rd\t\,X_4})
 \end{gathered}
 \label{e:tA}
\end{equation}
and the remaining thirteen node-raised/lowered versions of these.

Finally, since $\IX=(\f_1|\j_I|F_2,F_3,F_4)$ is the worldline reduction of the supermultiplet $i(\bs\L{-}\ba{\bs\L})$ of regular component fields, Theorem~\ref{T:1} guarantees that its twisted variant,
\begin{equation}
  \Tw\IX=(\f_1^{\sss K}|\j^{\sss K}_I|F_2^{\sss K},F_3^{\sss K},F_4^{\sss K})
  \overset{\sss\text{(\ref{e:X>F})}}\longleftrightarrow
  (\x_2|X_3,\Dt{x}_1,X_4,\Dt{x}_2|\Dt\x_4,\Dt\x_1,\Dt\x_3)
 \label{e:TwX}
\end{equation}
is indeed the worldline reduction of a supermultiplet in a higher dimensional spacetime (such as the 1+3-dimensional one), the component fields of which are all ghosts: $\f_1^{\sss K}$ and $F_3^{\sss K}$ are anticommuting scalars, $\Dt{\f}_2^{\sss K}$ and $F_4^{\sss K}$ anticommuting pseudo-scalars, and $\j^{\sss K}_I$ are commuting spinors. The $Q$-action amongst these ghost component fields is identical to the one within the supermultiplet $(\x_2|X_3,\Dt{x}_1,X_4,\Dt{x}_2|\Dt\x_1,\Dt\x_4,\Dt\x_3)$, which has in turn been manifestly obtained from the twisted-chiral supermultiplet depicted on the right-hand side of\eq{e:CtCA} by means of node-raising. Note that:
\begin{enumerate}\itemsep=-3pt\vspace*{-3mm}
  \item $(\x_2|X_3,\Dt{x}_1,X_4,\Dt{x}_2|\Dt\x_1,\Dt\x_4,\Dt\x_3)$ can extend to a worldsheet supermultiplet, with the Lorentz group $\Spin(1,1)$. However, owing to the obstruction described in Ref.\cite{rGH-obs}, this can only be a supermultiplet of worldsheet $(1,3)$- or $(3,1)$-supersymmetry\cite{rH-WWS}.
  \item $(\x_2|X_3,\Dt{x}_1,X_4,\Dt{x}_2|\Dt\x_1,\Dt\x_4,\Dt\x_3)$ cannot extend to a $(d\,{>}\,2)$-dimensional spacetime: already for $d=3$, the irreducible spinors of $\Spin(1,2)$ have two components, and it is impossible to Lorentz-covariantly separate the spinor component $\x_2$ from the $\Dt\x_1,\Dt\x_4,\Dt\x_3$.
  \item In turn, the ghost supermultiplet $(\f_1^{\sss K}|\j^{\sss K}_I|F_3^{\sss K},\Dt{\f}_2^{\sss K},F_4^{\sss K})$ straightforwardly extends to 1{+}2- and also to 1{+}3-dimensional spacetime.
  Identifying $\IY$ with the gauge vector supermultiplet in the Wess-Zumino gauge, $\f_1^{\sss K}\in\Tw\IX$ and its conjugate momentum would correspond to the usual Faddeev-Popov-De~Witt ghosts, the rest of $\Tw\IX$ to their off-shell supersymmetry completion; see\eq{e:CtCg} below.
\end{enumerate}\vspace*{-3mm}

 Similarly, $\IY=(\6({-}x_1),x_3,x_2|{-}\x_1,\x_2,\x_3,\x_4|X_4)$ is the worldline reduction of the supermultiplet $\{\IV\simeq\IV{+}i(\bs\L{-}\ba{\bs\L})\}$ of regular component fields in the Wess-Zumino gauge, where the component fields of $i(\bs\L{-}\ba{\bs\L})$ are used to cancel component fields in $\IV$.
 Then,
\begin{equation}
  \Tw\IY={}\big(({-}x_1^{\sss K}),x_3^{\sss K}),x_2^{\sss K}
          \big|{-}\x^{\sss K}_1,\x^{\sss K}_2,\x^{\sss K}_3,\x^{\sss K}_4
           \big|X_4^{\sss K}\big)
   \overset{\text{(\ref{e:X>F})}}\longleftrightarrow
  \big({-}\j_2,\j_1,\j_4\big|{-}F_3,\Dt\f_1,F_4,\Dt\f_2\big|\Dt\j_3\big)
\end{equation}
showing the result of a subsequent application of $\vdt$ on every component field in the right-hand side supermultiplet. Again, $\Tw\IY$ is the worldline reduction of a supermultiplet in a higher dimensional spacetime, the component fields of which are all ghosts. Amongst these, the $Q$-action is identical to the one in the supermultiplet $({-}\j_2,\j_1,\j_4|{-}F_3,\Dt\f_1,F_4,\Dt\f_2|\Dt\j_3)$, which has in turn been manifestly obtained from the chiral supermultiplet depicted on the left-hand side of\eq{e:CtCA} by means of node-raising. In turn, the regular (non-ghost) supermultiplet $({-}\j_2,\j_1,\j_4|{-}F_3,\Dt\f_1,F_4,\Dt\f_2|\Dt\j_3)$ cannot extend to spacetimes beyond $d\,{=}\,2$ since $\Dt\j_3$ can be separated from $\j_1,\j_2,\j_4$ in a Lorentz-covariant way only for $d\leqslant2$. In fact, even on the worldsheet, this can only be a supermultiplet for $(3,1)$- or $(1,3)$-supersymmetry\cite{rH-WWS,rGH-obs}.

\section{Off-Shell Supersymmetric BRST and BV Frameworks}
\label{s:G}
Fields which have the ``opposite'' spin-statistics assignment appear in covariant treatments of gauge symmetry and other systems with constraints. For Yang-Mills type of gauge symmetry, the BRST treatment suffices in all generality; see the texts~\cite{rSW2,rWS-Fields,rMS-QFT,rEZ-QFT3}. For other systems of constraints that are linearly dependent or do not close as an algebra, the more general Batalin-Vilkovisky framework is required\cite{rEZ-QFT3}, wherein field-variables with the ``opposite'' spin-statistics assignment are called antifields; see Chapter~XII of Ref.\cite{rWS-Fields}. For the sake of simplicity and uniformity, I will call these ``ghosts'' as well and restrict the discussion here to simple $(N\,{=}\,4)$ supersymmetry in 4-dimensional spacetime unless otherwise stated.

\paragraph{Ghost Supermultiplets:}
In models that need ghosts but also exhibit supersymmetry, ghosts have superpartners and so form supermultiplets. Indeed, the well-known (and manifestly supersymmetric) superfield method has been used already by 1983 to include Faddeev-Popov-De~Witt ghosts, as presented in the very first textbook on supersymmetry\cite{r1001}. While a consistent and fully off-shell dynamical treatment is not known for arbitrarily many supersymmetries, the ghost supermultiplets will remain indispensable since they are already for $N\,{=}\,4$.

 In particular, for simply ($N\,{=}\,4$) supersymmetric Yang-Mills gauge theories in 4-dimensional spacetime, the real Faddeev-Popov-De~Witt ghost fields $c_1$ and $c'_1$ are incorporated as the real parts of the lowest component fields of superfields \Bc, $\Bc'$ and their conjugates: $c_1=\Ree(\Bc|)$ and $c'_1=\Ree(\Bc'|)$.
 These superfields satisfy the ``chiral'' (and conjugate) superdifferential conditions:
\begin{equation}
  \bDb_{\dt\a}\,\Bc = 0 = \bD_{\a}\,\ba{\Bc}
   \qquad\text{and}\qquad
  \bDb_{\dt\a}\,\Bc' = 0 = \bD_{\a}\,\ba{\Bc}'
 \label{e:ChiG}
\end{equation}
where $\a,\dt\a=1,2$ (or $+,-$ as in Eqs.\eq{e:SuSyDc}, if one prefers). Adhering to the same real basis choices as above\eq{e:CtCA} and\eq{e:SuSyDc1}, a standard chiral superfield and its ghost variant, say \Bc, are respectively depicted as follows:
\begin{equation}
 \vC{\begin{picture}(55,36)(0,0)
   \put(0,0){\includegraphics[width=50mm]{N4B242.pdf}}
     \put(-2,30){\Cx{\large$\F$:}}
     \put(-2,26){\Cx{(chiral)}}
     \put(15,32){\Rx{$F_3$}}
     \put(35,32){\Lx{$F_4$}}
     \put(1.5,17){\Rx{$\j_1$}}
     \put(15.5,17){\Rx{$\j_2$}}
     \put(34.5,17){\Lx{$\j_3$}}
     \put(48.5,17){\Lx{$\j_4$}}
     \put(15,2){\Rx{$\f_1$}}
     \put(35,2){\Lx{$\f_2$}}
 \end{picture}}
  \qquad\textit{vs.}\qquad\quad\qquad
 \vC{\begin{picture}(55,36)(-5,0)
    % \put(7,3.5){\C{13}{\oval(29.0,7.0)}}
   \put(0,0){\includegraphics[width=50mm]{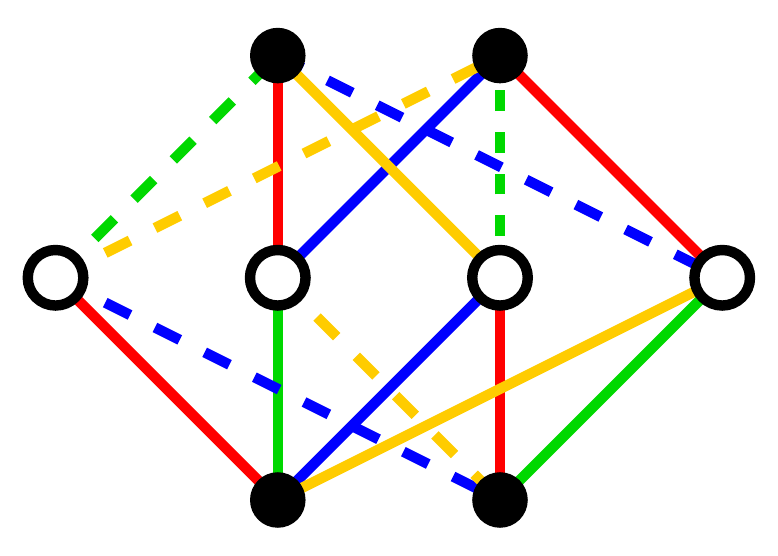}}
     \put(-2,30){\Cx{\large{\bf C}:}}
     \put(-6,26){\Cx{(twisted-chiral)}}
     \put(15,32){\Rx{$C_3$}}
     \put(35,32){\Lx{$C_4$}}
     \put(1.5,17){\Rx{$\c_1$}}
     \put(15.5,17){\Rx{$\c_2$}}
     \put(34.75,17){\Lx{$\c_3$}}
     \put(49,17){\Lx{$\c_4$}}
     \put(15,2){\Rx{{\footnotesize FPDW ghost\,=\,}$c_1$}}
     \put(35,2){\Lx{$c_2$}}
 \end{picture}}
 \label{e:CtCg}
\end{equation}
In Ref.\cite{r1001}, both of the supermultiplets \Bc, $\Bc'$ and their conjugates are all referred to as ``chiral scalars'', since they all satisfy the same superdifferential conditions\eq{e:ChiG} as does a chiral superfield\eq{e:C}. However, the straightforward computation of $\CP_{\sss\{\C31,\C12,\C63,\C84\}}(\F)=+1$ {\em\/vs\/}.\ $\CP_{\sss\{\C31,\C12,\C63,\C84\}}(\Bc)=-1$ proves that within the ghost superfields \Bc\ and $\ba{\Bc}$, supersymmetry acts in the manner of a twisted chiral supermultiplet\eq{e:tC}, rather than in the manner of a chiral one\eq{e:C}.

Indeed, upon dimensionally reducing the 4-dimensional superfields $\F$ and \Bc\ to the worldline, the procedure\eq{e:X>FA} applies verbatim:
\begin{equation}
 \vC{\begin{picture}(160,32)(0,-1)
   \put(0,0){\includegraphics[width=160mm]{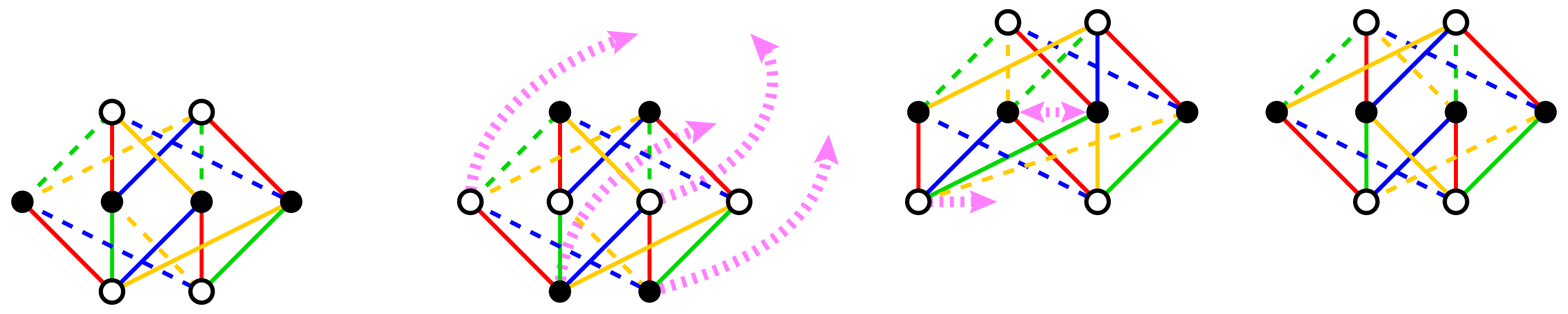}}
     \put(1,3){$\F$:}
     \put(47,3){\Bc:}
     \put(73.5,19){$\SSS\dt{c}_1$}
     \put(83,19){$\SSS\dt{c}_2$}
     \put(65,28){$\SSS\dt\c_1$}
     \put(73.5,28){$\SSS\dt\c_3$}
     \put(134.5,10){$\SSS\c_2$}
     \put(150.5,10){$\SSS\c_4$}
     \put(126,19){$\SSS C_3$}
     \put(135,19){$\SSS\dot{c}_1$}
     \put(150,19){$\SSS C_4$}
     \put(159.5,19){$\SSS\dt{c}_2$}
     \put(134.5,29){$\SSS\dot\c_1$}
     \put(150.5,29){$\SSS\dot\c_3$}
     \put(16,29){\Cx{\small chiral}}
     \put(16,25){\Cx{\footnotesize left-hand side of\eq{e:CtCA}}}
     \put(38,10.5){\Cx{\it vs.}}
     \put(144,5){\Cx{\small twisted-chiral}}
     \put(144,1){\Cx{\footnotesize right-hand side of\eq{e:CtCA}}}
 \end{picture}}
 \label{e:C>gC}
\end{equation}
Comparison with the pair\eq{e:CtCA} then reconfirms the ghost superfields \Bc\ and $\Bc'$\eq{e:ChiG} to indeed have the {\em\/twisted\/} chiral supermultiplet structure, while providing an off-shell supersymmetric completion for the (real) Faddeev-Popov-De~Witt ghost field $c_1$ and antighost field $c'_1$, respectively.

The computation of $\CP_\bc(\fA)$ however requires neither a worldline dimensional reduction nor the transformation\eq{e:C>gC}$\,{\approx}\,$(\ref{e:X>FA}), the inverse of which is non-local even on the worldline.
 This feature guarantees the usability of $\CP_\bc(\fA)$ to all higher-dimensional supersymmetry.
 Indeed, $\CP_\bc(\Bc)=-1=-\CP_\bc(\F)$ determines unambiguously that the ghost supermultiplet \Bc\ depicted to the right in\eq{e:CtCg} is a twisted variant of the chiral $\F$ depicted on the right-hand side. Also, the combination $i(\ba\Bc{-}\Bc)$ that appears in the BRST transformation\eq{e:BRST} below has precisely the supersymmetry structure\eq{e:TwX} guaranteed by Theorem~\ref{T:1}.

Finally, the BRST formalism does not suffice in theories with open or reducible gauge symmetries. Its appropriate generalization for these much more general cases was found first by Zinn-Justin and was then completed by Batalin and Vilkovisky\cite{rSW2}. In this formalism, to every field $\f$ one introduces a corresponding antifield $\f^\ddag$, which differs from $\f$ only in that $\f^\ddag$ is anticommuting if $\f$ is commuting, and {\em\/vice versa\/}. In supersymmetric field theories then, antifields must combine into supermultiplets, much as regular fields do. Since the statistics of antifields is opposite of that of the original fields, antifield supermultiplets will again furnish the twisted variant of the supersymmetry representation furnished by the corresponding fields, as per Theorem~\ref{T:1} and Corollary~\ref{C:1}.

\paragraph{Off-Shell Dynamics:}
The fully off-shell formulation of supersymmetric Yang-Mills models requires adding to the so-called vector superfield $\IV=\ba\IV$ the conjugate pair of Faddeev-Popov-De~Witt ghost superfields\eq{e:ChiG} as well as a Nakanishi-Lautrup chiral superfield $\IB$ and its conjugate $\ba\IB$, satisfying $\bDb_{\dt\a}\IB=0=\bD_\a\ba\IB$.
 With these, the original gauge transformation is made nilpotent by replacing the gauge parameter superfield $\bs\L\to i\e\mkern1.5mu\Bc$ and $\d_{\sss\text{YM}}\to\d_{\sss\text{BRST}}\Defl\e\BRST$, where $\e$ is a real anticommuting constant.
 Then,
\begin{subequations}
 \label{e:BRST}
 \begin{alignat}9
 \BRST\IV &= \Bc-\ba\Bc
  \qquad&\&&\qquad
 \BRST\Bc &= 0 &&= \BRST\ba\Bc,
 \label{e:BRSTc}\\
 \BRST\Bc' &=-i\,\IB,\quad \BRST\ba\Bc'=-i\,\ba\IB,
  \qquad&\&&\qquad
 \BRST\IB &= 0 &&= \BRST\ba\IB,
 \label{e:BRSTc'}
\end{alignat}
\end{subequations}
is fully off-shell nilpotent, $\BRST^2=0$, as necessary for consistent quantum theory.

The superfield Lagrangian terms involving the ghost superfields $\Bc,\ba\Bc,\Bc'\ba\Bc'$ and the Nakanishi-Lautrup superfields $\IB,\ba\IB$ may be rewritten in terms of the real combinations $i(\ba\Bc{-}\Bc)$, $(\Bc'{+}\ba\Bc')$ and $(\IB{+}\ba\IB)$. 
 Together with $\IV$, these then provide off-shell degrees of freedom\ft{While a chiral superfield $\F$ has $(2|4|2)$ independent real component field degrees of freedom in ascending mass-dimensions, for the real combination $\F{+}\ba\F$ this count is $(1|4|3)$: The lowest component $(\F{+}\ba\F)|=\Ree(\f)$ is simply the real part of the lowest component $\f=\F|$, the imaginary part of which shows up as $\vd_\m\!\Imm(\f)\in(\F{+}\ba\F)$, with the mass-dimension $[\f]{+}1$; the analogous holds for $(\F{-}\ba\F)$\cite{r1001}; see appendix.} at various mass-dimensions, tabulated here for $(d\,{\leqslant}\,4)$-dimensional spacetime:
\begin{equation}
  \begin{array}{@{} r|rrrrrrrl @{}}
  & \bs{\frc{d-4}2} & \bs{\frc{d-3}2} & \bs{\frc{d-2}2} & \bs{\frc{d-1}2}
  & \bs{\frc{d}2}   & \bs{\frc{d+1}2} & \bs{\frc{d+2}2} & \text{\bsf type} \\ 
    \toprule
 \IV            & 1  & 4  & 6  & 4 & 1  & 0  & 0   & \text{regular}\\ 
 \Bc{-}\ba\Bc   & -1 & -4 & -3 & 0 & 0  & 0  & 0   & \text{ghost} \\ 
 \Bc'{+}\ba\Bc' & 0  & 0  & 0  & 0 & -1 & -4 & -3  & \text{ghost} \\ 
 \IB{+}\ba\IB   & 0  & 0  & 0  & 0 &  1 &  4 &  3  & \text{regular} \\ 
    \midrule
\text{\bsf net:}& 0  & 0  & 3  & 4 & 1  & 0  & 0   & \text{regular} \\
    \bottomrule
  \end{array}
\end{equation}
Ghost fields are counted here as negative regular degrees of freedom since ghost fields effectively subtract the contributions of regular fields in the computation of every correlation function.

The final, net count of regular degrees of freedom is then identical to the one in $\IV$ when reduced in the Wess-Zumino gauge. However, the two schemes differ significantly:
\begin{enumerate}\itemsep=-3pt\vspace{-2mm}
 \item Models where $\IV$ is reduced in the Wess-Zumino gauge, the component gauge vector field $A_\m$ is subject to the gauge equivalence, $A_\m\simeq A_\m{-}\vd_\m\l$, and so is not fully off shell: integration over $A_\m$ in the path integral must be restricted, \eg, to the $\vd^\m A_\m=0$ gauge-slice. Also, the imposition of the Wess-Zumino gauge breaks supersymmetry, and {\em\/superymmetric quantization in the Wess-Zumino gauge is impossible\/}\cite{r1001}.
 \item Models using the full complement of superfields $\{\IV|(\Bc{-}\ba\Bc),(\Bc'{+}\ba\Bc')|(\IB{+}\ba\IB)\}$ leave all component fields unrestricted by any spacetime differential equation (or other gauge condition), \ie, fully off shell, and suitable for path-integral integration. Desired gauge conditions are incorporated within BRST-exact terms in the action, so as to ease particular computations while leaving physical observables unaffected.
\end{enumerate}

\section{Conclusions}
\label{s:C}
The original supermultiplet twisting\cite{rTwSJG0} (see also Ref.\cite{rGHR,rHSS,r6-1}) is well-defined only in $d\leqslant2$ dimensional spacetimes. An alternate supermultiplet twisting Definition~\ref{D:Tw} is given herein, which is unique in that it is valid in spacetimes of all dimensions and coincides with the original in $d\leqslant2$ spacetime where both are valid. This is proven both by means of the non-local field redefinitions\eq{e:X>F} valid in $d\leqslant2$ spacetimes as well as using the cycle parity invariant (Definition~\ref{D:CP}), which is valid in all spacetimes.

 When non-trivial (see Ref.\cite[ and references therein]{r6-3.1}), this general definition of off-shell supermultiplet twisting maps regular (non-ghost) supermultiplets to ghost supermultiplets and {\em\/vice versa\/}.
  This conclusion may be avoided only in spacetimes of low enough dimension, \ie, on the worldsheet and the worldline, where the Lorentz group $\Spin(1,d{-}1)$ is abelian, the irreducible tensorial and spinorial representations are all 1-dimensional and interchangeable at will.
 
All proofs are presented for supermultiplets depictable by adinkras, where each supercharge maps each component field into precisely one other component field or its spacetime derivative.
 However, the main results apply much more generally upon observing that the superdifferential constraint and equivalence construction\eqs{e:DU=0}{e:U/DU} can be employed iteratively {\em\/and indefinitely\/}, to construct ever more complex (and larger) off-shell supermultiplets, to all of which the twisting in Definition~\ref{D:Tw}, Theorem~\ref{T:1} and its Corollary~\ref{C:1} apply. Given the initial indications in Refs.\cite{rTHGK12,rGHHS-CLS,rGIKT12}, virtually none of these supermultiplets are adinkraic. In turn, it is not known if there exist off-shell supermultiplets that cannot be constructed in this way.

In generalizations of the sequence\eq{e:XYZ} such as studied in Ref.\cite{rTHGK12}, $\IX$ and $\IY$ are adinkraic only in exceptional cases. However, as long as the central supermultiplet $\IU$ is a direct sum of intact supermultiplets, it will follow that $\CP_\bc(\IY)=-\CP_\bc(\IX)$. This prompts:
\begin{conj}\label{C:2}
Given a short exact sequence\ft{That $\IX\overset\a\into\IU\overset\w\onto\IY$ is a ``short exact sequence'' means that: $\a$ is a 1--1 injection, $\w$ is a surjection and $\w\circ\a=0$, so that
  $(\IX\,{\simeq}\,\im(\a)\,{\subset}\,\IU)\,{=}\,\ker(\w)$ and
 $\IY\,{=}\,\cok(\a)\,{=}\,\{\IU/\im(\a)\}$.}
 $\IX\overset\a\into\IU\overset\w\onto\IY$ of strict homomorphisms of off-shell supermultiplets\cite{r6-3.2} between off-shell supermultiplets, \ie, given $\IX\,{:}~\w(\a\6(\IX))=0$, $\a(\IX)\approx\IX$ and $\IY\,{:=}\,\{\IU\simeq\IU{+}\a(\IX)\}$, 
\begin{equation}
  \CP_\bc(\IX)+\CP_\bc(\IY) = \CP_\bc(\IU)
\end{equation}
for every fixed cycle \bc\ of distinct edge-colors, for all off-shell supermultiplets $\IX,\IU,\IY$ of $N$-extended worldline supersymmetry with no central extension.
\end{conj}

While for special Lie groups---such as $\Spin(8)$---the tensorial and spinorial representations are interchangeable without being 1-dimensional, there exists no $\Spin(1,d{-}1)$ Lorentz group with this property. As observed in Ref.\cite{rFGH}, the representations spanned by the bosons and the fermions are additionally restricted by the requirements
\begin{equation}
  \sR_L(Q)\otimes\sR_L(\f)\supset\sR_L(\j)
   \qquad\text{and}\qquad
  \sR_L(Q)\otimes\sR_L(\j)\supset\sR_L(\f),
\end{equation}
where $\sR_L(Q)$ for our present purposes denotes the representation of the Lorentz group spanned by the supercharges, while $\sR_L(\f)$ and $\sR_L(\j)$ denote the Lorentz representation spanned by the bosonic and fermionic component fields, respectively. For the regular spin-statistics correspondence, $\sR_L(\f)$ must be a tensorial representation while $\sR_L(\j)$ must be a spinorial one; for the ``wrong'' spin-statistics correspondence in ghost supermultiplets, this is reversed.

Now, the Haag-\Lv opusa\'nski-Sohnius theorem guarantees that $\sR_L(Q)$ must in all circumstances be Lorentz spinors with anticommuting components $Q_I$. It then follows that indecomposable supermultiplets consist of:
\begin{enumerate}\itemsep=-3pt\vspace{-2mm}
 \item either regular (commuting bosonic and anticommuting fermionic) components fields,
 \item or ghost (anticommuting bosonic and commuting fermionic) components fields.
\end{enumerate}
The nonlinear sigma-models of Ref.\cite{rGHR} require otherwise identical regular supermultiplets of both un-twisted and twisted kind, and so cannot be generalized to $(d\,{\geq}\,4)$-dimensional spacetimes. Nonlinear sigma-models in $(d\,{\geq}\,4)$-dimensional spacetimes thus cannot have non-K\"ahler target spaces \'a la Gates, Hull and Ro\v{c}ek\cite{rGHR}, confirming the so-called Zumino theorem.

Finally, Yang-Mills models with simple ($N\,{=}\,4$) supersymmetry in four-dimensional spacetime\cite{rWS-Fields} include the familiar Faddeev-Popov-De~Witt ghosts within supermultiplets that obey the familiar chiral superfield defining conditions, but---as per Theorem~\ref{T:1}---in fact have the twisted chiral supermultiplet structure. This is most easily proven by computing the cycle parity from Definition~\ref{D:CP}. This result clearly generalizes to the supersymmetric Zinn-Justin-Batalin-Vilkovisky treatment of other constrained systems.
 While completely off-shell treatments of general supersymmetric systems with constraints (incl.\ gauge theories) in models with $N\,{\geqslant}\,8$ supersymmetries are not known, they will have to include ghost supermultiplets (and the above results) since all the known $(N\,{\leqslant}\,8)$-supersymmetric cases do, and they are included in the higher-dimensional/supersymmetry ones by dimensional/supersymmetry reduction.

\bigskip\paragraph{\bfseries Acknowledgments:}
 Many thanks to Mr.~James Parker for the advance copy of Ref.\cite{rJP-DBt}, written under the mentorship of Prof.~G.D.~Landweber,
  to S.J.~Gates, Jr.\ and W.~Siegel for helpful discussions
  and to the Referee for prompting the material of Section~\ref{s:G}.
 I am grateful to the Department of Energy for the generous support through the grant DE-FG02-94ER-40854, as well as the Physics Department of the Faculty of Natural Sciences of the University of Novi Sad, Serbia, for recurring hospitality and resources.

\appendix
\section*{Appendix}
The standard definition of a chiral superfield, rewritten in the particular basis choice\eq{e:SuSyDc}:
\begin{subequations}
% \label{e:}
\begin{equation}
  \bDb_\pm\F=0,\qquad
  [\C3{\rD_1}-i\C6{\rD_3}](\IA+i\IB)=0=[\C1{\rD_2}-i\C8{\rD_4}](\IA+i\IB),
\end{equation}
implies, by separating the real and imaginary parts, the four equations\\[-2mm]
\begin{minipage}{.48\hsize}
\begin{align}
 \C3{\rD_1}\IA+\C6{\rD_3}\IB&=0, \label{e:1A+3B}\\
 \C6{\rD_3}\IA-\C3{\rD_1}\IB&=0, \label{e:3A-1B}
% \label{e:}
\end{align}
\end{minipage}\hfill
\begin{minipage}{.48\hsize}
\begin{align}
 \C1{\rD_2}\IA+\C8{\rD_4}\IB&=0, \label{e:2A+4B}\\
 \C8{\rD_4}\IA-\C1{\rD_2}\IB&=0. \label{e:4A-2B}
% \label{e:}
\end{align}
\end{minipage}
\end{subequations}
\\[3mm]
Applying judicious superderivatives on these short equations and combining them, we obtain:
\begin{alignat}9
 \C3{\rD_1}\C1{\rD_2}\IA
 &\isBy{\sss(\ref{e:2A+4B})}-\C3{\rD_1}\C8{\rD_4}\IB&
 &= \C8{\rD_4}\C3{\rD_1}\IB&
 &\isBy{\sss(\ref{e:3A-1B})} \C8{\rD_4}\C6{\rD_3}\IA&
 &~= -\C6{\rD_3}\C8{\rD_4}\IA;\\[2mm]
 \C3{\rD_1}\C6{\rD_3}\IA
 &\isBy{\sss(\ref{e:3A-1B})} \C3{\rD_1}\C3{\rD_1}\IB&
 &=H\IB&
 &~=\C1{\rD_2}\C1{\rD_2}\IB&
 &\isBy{\sss(\ref{e:4A-2B})}\C1{\rD_2}\C8{\rD_4}\IA; \label{e:13A=24A}\\[2mm]
 \C3{\rD_1}\C8{\rD_4}\IA
 &\isBy{\sss(\ref{e:4A-2B})}\C3{\rD_1}\C1{\rD_2}\IB&
 &=-\C1{\rD_2}\C3{\rD_1}\IB&
 &\isBy{\sss(\ref{e:3A-1B})}-\C1{\rD_2}\C6{\rD_3}\IA.
\intertext{Similarly,}
 \C3{\rD_1}\C1{\rD_2}\IB
 &\isBy{\sss(\ref{e:4A-2B})}\C3{\rD_1}\C8{\rD_4}\IA&
 &=-\C8{\rD_4}\C3{\rD_1}\IA&
 &\isBy{\sss(\ref{e:1A+3B})} \C8{\rD_4}\C6{\rD_3}\IB&
 &~= -\C6{\rD_3}\C8{\rD_4}\IB;\\[2mm]
 \C3{\rD_1}\C6{\rD_3}\IB
 &\isBy{\sss(\ref{e:1A+3B})}-\C3{\rD_1}\C3{\rD_1}\IA&
 &=-H\IA&
 &~=-\C1{\rD_2}\C1{\rD_2}\IA&
 &\isBy{\sss(\ref{e:2A+4B})}\C1{\rD_2}\C8{\rD_4}\IB;\\[2mm]
 \C3{\rD_1}\C8{\rD_4}\IB
 &\isBy{\sss(\ref{e:2A+4B})}-\C3{\rD_1}\C1{\rD_2}\IA&
 &=\C1{\rD_2}\C3{\rD_1}\IA&
 &\isBy{\sss(\ref{e:1A+3B})}-\C1{\rD_2}\C6{\rD_3}\IB.
% \label{e:}
\end{alignat}
Comparing the starting (far left-hand side) expression with the ending (far right-hand side) expression proves that both real and the imaginary parts of a chiral superfields $\IA=\Ree(\F)=\fc12(\F{+}\ba\F)$ and $\IB=\Imm(\F)=\fc1{2i}(\F{-}\ba\F)$ satisfy the same (chiral) ``self-duality'' conditions:
\begin{alignat}9
 \ID^+_{IJ}\IA&=0:\qquad
 \C3{\rD_1}\C1{\rD_2}\IA &= -\C6{\rD_3}\C8{\rD_4}\IA,&\quad
 \C3{\rD_1}\C6{\rD_3}\IA &= +\C1{\rD_2}\C8{\rD_4}\IA,&\quad
 \C3{\rD_1}\C8{\rD_4}\IA &= -\C1{\rD_2}\C6{\rD_3}\IA;\\[2mm]
 \ID^+_{IJ}\IB&=0:\qquad
 \C3{\rD_1}\C1{\rD_2}\IB &= -\C6{\rD_3}\C8{\rD_4}\IB,&\quad
 \C3{\rD_1}\C6{\rD_3}\IB &= +\C1{\rD_2}\C8{\rD_4}\IB,&\quad
 \C3{\rD_1}\C8{\rD_4}\IB &= -\C1{\rD_2}\C6{\rD_3}\IB.
% \label{e:}
\end{alignat}
In turn, the same Adinkra depicts them both faithfully\cite{r6-1.2}; for example,
\begin{equation}
 \vC{\begin{picture}(150,30)(-5,-2)
       \put(23.75,0){\includegraphics[height=28mm]{N4B143+x.pdf}}
        \put(5,25){$\IA$:}
        \put(37,25){\cB{$\C3{\rD_1}\C1{\rD_2}\IA|$}}
        \put(75,25){\cB{$\C3{\rD_1}\C6{\rD_3}\IA|$}}
        \put(113,25){\cB{$\C3{\rD_1}\C8{\rD_4}\IA|$}}
        \put(28,10){\cB{$\C3{\rD_1}\IA|$}}
        \put(60,10){\cB{$\C1{\rD_2}\IA|$}}
        \put(90,10){\cB{$\C6{\rD_3}\IA|$}}
        \put(122,10){\cB{$\C8{\rD_4}\IA|$}}
        \put(75,0){\cB{$\IA|$}}
     \end{picture}}
 \label{e:IA}
\end{equation}
It is easy to verify that $\CP_{\{\C31,\C12,\C63,\C84\}}(\IA)=+1$, proving that both $\IA$ and $\IB$ have the chiral chromo\-to\-po\-logy---although they are not locally equivalent to the standard chiral superfield $\F$, depicted on the left-hand side of\eq{e:CtCA}.

Inverting\eq{e:SuSyDc1}, we have
\begin{equation}
 \C3{\rD_1}=\fc12[\bD_++\bDb_+],\quad
 \C1{\rD_2}=\fc12[\bD_-+\bDb_-],\quad
 \C6{\rD_3}=\fc1{2i}[\bD_+-\bDb_+],\quad
 \C8{\rD_4}=\fc1{2i}[\bD_--\bDb_-].
% \label{e:}
\end{equation}
The bosonic component fields appearing in the Adinkra\eq{e:IA} may also be defined (suitable for worldsheet calculations) as
\begin{alignat}9
 \IA|
 &= \fc12(\F{+}\ba\F)| = \Re(\f);\\[1mm]
 \C3{\rD_1}\C1{\rD_2}\IA|
 &= \fc12[\bD_++\bDb_+]\fc12[\bD_-+\bDb_-]\fc12(\F{+}\ba\F)|
  = \fc18[\bD_++\bDb_+](\bD_-\F + \bDb_-\ba\F)|\nn\\
 &= \fc18([\bD_+\bD_-\F + \bDb_+\bDb_-\ba\F)|
  = \fc12(F{+}\ba{F}) = \Ree(F);\\[1mm]
 \C3{\rD_1}\C6{\rD_3}\IA|
 &= \fc12[\bD_++\bDb_+]\fc1{2i}[\bD_+-\bDb_+]\fc12(\F{+}\ba\F)|
  = \fc1{8i}[\bD_++\bDb_+](\bD_+\F - \bDb_+\ba\F)|\nn\\
 \vC{$\SSS(\ref{e:13A=24A})$}\|~~\quad
 &= \fc1{8i}([2i\vd_\pp\F - 2i\vd_\pp\ba\F)|
  = \fc14\vd_\pp(\f{-}\ba\f) = \fc{i}2(\vd_\pp\Imm\f); \label{e:++Imf}\\[0mm]
 \C1{\rD_2}\C8{\rD_4}\IA|
 &= \fc12[\bD_+-\bDb_-]\fc1{2i}[\bD_--\bDb_-]\fc12(\F{+}\ba\F)|
  = \fc1{8i}[\bD_-+\bDb_-](\bD_-\F - \bDb_-\ba\F)|\nn\\
 &= \fc1{8i}([2i\vd_\mm\F - 2i\vd_\mm\ba\F)|
  = \fc14\vd_\mm(\f{-}\ba\f) = \fc{i}2(\vd_\mm\Imm\f); \label{e:--Imf}\\[1mm]
 \C3{\rD_1}\C8{\rD_4}\IA|
 &= \fc12[\bD_++\bDb_+]\fc1{2i}[\bD_--\bDb_-]\fc12(\F{+}\ba\F)|
  = \fc1{8i}[\bD_++\bDb_+](\bD_-\F - \bDb_-\ba\F)|\nn\\
 &= \fc1{8i}([\bD_+\bD_-\F - \bDb_+\bDb_-\ba\F)|
  = \fc12(F{-}\ba{F}) = \Imm(F).
% \label{e:}
\end{alignat}
Comparing Eqs.\eq{e:++Imf} with\eq{e:--Imf} implies that
\begin{equation}
 \fc{i}2(\vd_\pp\Imm\f) \isBy{\sss(\ref{e:13A=24A})} \fc{i}2(\vd_\mm\Imm\f),
% \label{e:}
\end{equation}
which can hold only on the worldline embedded within the worldsheet, where $\vd_\pp=\vd_\mm$, \ie, where the fields are independent of the spatial worldsheet coordinate. This conclusion agrees with the ``bow-tie theorem'' of Ref.\cite{rGH-obs}.

We thus conclude that the Adinkra\eq{e:IA} depicts an irreducible and indecomposable off-shell superfield only on the worldline. For supersymmetries in all other nontrivial spacetimes this Adinkra depicts both the ``real'' and the ``imaginary'' linear combinations, $(\F{+}\ba\F)$ and $i(\ba\F{-}\F)$ respectively, of an off-shell chiral superfield and its Hermitian conjugate.

%\begingroup
%\bibliographystyle{elsart-numX}
%\raggedright
%\bibliography{Refs}
%\endgroup

\end{document}